\documentclass[a4paper,11pt]{article}
\usepackage{amsmath}
\usepackage{amsfonts}
\usepackage{graphicx}
\usepackage{amssymb}
\usepackage{epsfig}
\usepackage{empheq}

\usepackage{hyperref}
\usepackage{ulem}

\def\ba{\begin{eqnarray}}
\def\ea{\end{eqnarray}}
\def\be{\begin{equation}}
\def\ee{\end{equation}}

\def\nn{\nonumber}

\def\O{\mathcal{O}}

\def\eps{\epsilon}
\def\L{\mathcal{L}}

\def\d{\partial}

\def\phis{\phi_0^\star}

\def\Ohh{\mathcal{O}\left(\hbar^2\right)}

\usepackage{tikz}
\usetikzlibrary{arrows,shapes}
\usetikzlibrary{trees}
\usetikzlibrary{matrix,arrows} 				% For commutative diagram
\usetikzlibrary{positioning}				% For "above of=" commands
\usetikzlibrary{calc,through}				% For coordinates
\usetikzlibrary{decorations.pathreplacing}  % For curly braces
\usepackage{pgffor}							% For repeating patterns

\usetikzlibrary{decorations.pathmorphing}	% For Feynman Diagrams
\usetikzlibrary{decorations.markings}

\numberwithin{equation}{section}

\begin{document}

\pagenumbering{Alph}

\begin{titlepage}

\vskip 1.5cm

\vskip 1.5cm

\begin{center}
{\LARGE \bf{Effective Potential in Finite Formulation of QFT}}
\vskip 1.5cm

\large{{\bf Sander Mooij and Mikhail Shaposhnikov}}
\vskip 1cm

Institute of Physics,  \'Ecole Polytechnique F\'ed\'erale de Lausanne, \\CH-1015 Lausanne, Switzerland

\vskip 1cm

\tt{sander.mooij@epfl.ch,~mikhail.shaposhnikov@epfl.ch}

\end{center}

\vskip1cm

\begin{center}
\large{{\bf Abstract}}
\end{center}
\vskip 0.5cm

In recent works \cite{Mooij:2021ojy, Mooij:2021lbc}, we have shown how $n$-point correlation functions in perturbative QFT can be computed without running into intermediate divergences. Here we want to illustrate explicitly that one can calculate the quantum effective potential by the same method. As a main example, we consider a theory with two fields having large and small vacuum expectation values (vev). We show that no fine-tuning between the {\it physical quantities} is needed to keep the hierarchy between the vevs of different fields.

\thispagestyle{empty}

\end{titlepage}

\pagenumbering{arabic}

\newpage
\section{Introduction}
\label{intro}
In recent works \cite{Mooij:2021ojy, Mooij:2021lbc}, we have shown how to compute $n$-point Green's functions in scalar field theories without running into large or infinite quantum loop corrections that need to be carefully adjusted in theories with several distinct mass scales. We have argued that the existence of such a finite QFT formalism indicates that the so-called ``hierarchy problem" is method-dependent and that no fine-tunings are needed between {\it physical parameters} even if the theory contains light and superheavy particles.  In this sense, the physics at the electroweak scale does not, in general, feel the physics of a UV completion living at much higher energy scales \cite{Appelquist:1974tg}, with notable exceptions related to experimentally challenged attempts to describe the Higgs boson in terms of a composite bound state, or within the theories with dynamical supersymmetry breaking, or in theories with large extra dimensions and low fundamental Planck scale (the updated reviews can be found in \cite{ParticleDataGroup:2022pth}). 

Our work \cite{Mooij:2021ojy} finds its place in a long debate about the meaning of infinite and ``finite but large" contributions to correlation functions. As for the infinities, the discussion goes back to the days of Dirac \cite{Salam:1987in} up to the present day, see for example the recent work \cite{Branchina:2022jqc}. Quadratically divergent quantum corrections in theories with fundamental scalar fields have been the major driving force of overwhelming literature on naturalness issues. A contemporary overview can be found, for example in \cite{Manohar:2018aog, Cohen:2019wxr}, plus references therein (and references in our paper \cite{Mooij:2021ojy}).

In \cite{Mooij:2021ojy, Mooij:2021lbc}, we mainly focussed on the computation of $n$-point correlation functions. This work aims to extend the formalism of finite QFT to the computation of the effective action of the theory, an instrument to uncover the ground state of the theories with spontaneous symmetry breaking \cite{Coleman:1973jx}. 

Historically, the gauge hierarchy problem has one of its roots in Grand Unified Theories (GUTs) of strong, weak and electromagnetic interactions \cite{Georgi:1974sy}. It is required that one of the GUT multiplets (say, $\underline{24}$ of SU(5)) has a very large vacuum expectation value (vev) of the order of the GUT scale $\mathcal{O}\left(10^{14} \right)$ GeV to break the symmetry down to that of the Standard Model, and another (say, $\underline{5}$ of SU(5)) has a vev defining the Fermi scale. 

In \cite{Gildener:1976ai} it was argued that the radiative corrections set up an upper a bound on a ratio of the vector boson masses of the order of ${\cal O}(1/\sqrt{\alpha})$, where $\alpha$ is the fine structure constant, making GUTs impossible. It was soon realised that this was not the case, and there is no limit to gauge hierarchies coming from perturbation theory \cite{Buras:1977yy, Weinberg:1978ym}. Still, it was reasoned that the hierarchy is not  ``natural'', as getting it would require fine-tunings between the different orders of the perturbation theory. ``Fine-tuning'' means in general that a specific combination of two large numbers (say, mass parameters of  $\underline{5}$ - and  $\underline{24}$ - plets of SU(5)) should be adjusted in such a way that it is very small. An equivalent reformulation of this problem is known as the ``doublet-triplet'' splitting problem (for a review see \cite{Mohapatra:1997sp} and for a recent reanalysis \cite{Boer:2019qkn}).  The SU(5) GUT symmetry breaking splits a fundamental 5-plet  into a light doublet (the Higgs field) and a scalar leptoquark - SU(3) triplet whose masses remain of the order of the GUT breaking scale. To keep this hierarchy the same type of fine-tuning is needed. 

Note that the gauge hierarchy (or doublet-triplet splitting) problem contains two layers. First, one needs an initial choice of parameters at the tree level to establish the hierarchy between the electroweak scale and the GUT scale. Second, one seemingly needs additional fine-tunings to keep this dramatic doublet-triplet splitting stable against loop corrections. 

The problem has been settled a long time ago.  First, the necessity of the initial fine-tuning depends on the choice of the ``fundamental parameters'' of the action and thus cannot be expressed in measurable physical terms. For instance, if the ``fundamental parameters'' are the mass terms of different GUT scalar multiplets, fine-tuning is needed. If the ``fundamental parameters'' are the vevs of the multiplets, no fine-tuning is necessary (see, e.g.  \cite{Boer:2019qkn}). As for the radiative corrections, they are not uniquely defined and depend on the renormalisation procedure. The latter can be constructed in such a way that no fine-tunings are needed in any order of perturbation theory \cite{Kazama:1981fx}. In other words, there is the physics problem ``why the GUT scale is so much larger than the Fermi scale'', but the problem about the stability of the hierarchy is artificial and related to the choice of formalism, but not the physics.

In this paper, we show how to arrive at the same conclusions within the finite QFT method that we used in \cite{Mooij:2021ojy, Mooij:2021lbc}.  We reframe this formalism in terms of the effective action. For definitiveness, we focus on the case of two interacting quartically coupled scalar fields with a large mass hierarchy in between them. We show how their effective action can be computed without the need of ``artificially" invoking large counter-terms that need to be balanced against large bare loop corrections to shield the physics of the light field from the influence of the heavy field. Moreover, we show how the choice of parameters at the tree level suffices to avoid those large quantum corrections blowing up the vev and the masses of the light field. In other words, loop corrections do not spoil the hierarchy imposed at the tree level. Consequently, there is no need for additional fine-tuning. Once again, the delicate, fine-tuned interplay between the physics of widely separated mass scales that {\it keeps} the Higgs vev small and the Higgs field light is just an artefact of the standard computational method. Precisely as we advocated in the previous paper, low-scale physics is insensitive to the way it is embedded in possible large-scale extensions.

The paper is organised as follows. In Section \ref{stapp} we set our notation and quickly review the standard approach to compute the effective potential. Section \ref{fin} is devoted to the generalisation of the finite Callan-Symanzik approach to computing correlation functions for the case of effective action. In Section \ref{twof}, we show how formalism can be used in a theory of two interacting scalar fields. Implications for the quantum stability of the electroweak scale are reserved for Section \ref{stable}. We conclude in Section \ref{conclu}.

As for our notational convenience, we choose to reinstall $\hbar$ in all equations, unless explicitly stated. We will only consider a one-loop approximation, as it is sufficient for demonstrating the main point of our study.

\section{Standard approach} 
\label{stapp}
In this section, we quickly review the standard approach \cite{Coleman:1973jx} to computing the renormalised effective action $\Gamma_{\rm eff}$ and effective potential $\Gamma$. We will restrict ourselves to the simple $\lambda \phi^4$ theory.  {We begin from the textbook definition of the effective action and specify renormalisation conditions. The final result is presented in its general form in  eq.~\ref{ve}. Its coefficients are given in (depending on the chosen renormalisation conditions) eqs.~\ref{vecoeff}, \ref{brocof} and \ref{brocof2}.}

We begin from the classical action\footnote{One could add a ``cosmological constant" $\Lambda$, which defines a zero-point function in eq.~\ref{clascorr}. However, since in QFT, there is no physical meaning in a constant term we do not discuss it here. We leave the quartically divergent vacuum bubble diagram (the fourth diagram in Fig.~\ref{gamdiag}) outside the discussion. Appendix \ref{cc} shows how such a cosmological constant could be included in the framework that we are about to present.} (in $-+++$ metric):
\be
\Gamma_{\rm cl} =-i\cdot  \int d^4 x \left[\frac{1}{2}\left(\d_\mu \phi_0\right)^2+\frac{m^2}{2}\phi_0^2 +\frac{\lambda}{4!}\phi_0^4 \right]\,.
 \label{classact}
\ee
The corresponding tree correlation functions are given by:
\be
\Gamma_{\rm cl}^{(2)} = i\left(k^2+m^2\right), \qquad \qquad \Gamma^{(4)}_{\rm cl} = -i\lambda\,.
\label{clascorr}
\ee
The quantum effective action  $\Gamma$ is the generating functional for the strongly connected Green's functions $\Gamma^{(n)}(x_1\dots x_n)$ and reads 
\be
\Gamma_{\rm eff} =\sum_n \frac{1}{n!}\int d^4x_1\dots d^4x_n ~\Gamma^{(n)}(x_1\dots x_n)\phi_0(x_1)\dots \phi_0(x_n)\,.
\label{pos}
\ee
Equivalently, the effective action can be written as an expansion in powers of derivatives:
\be
\Gamma_{\rm eff} = -i\int d^4 x \left[ \Gamma(\phi_0)+ \frac{1}{2}\left(\d_\mu \phi_0\right)^2 Z(\phi_0)+\dots   \right]\,.
 \label{derexp}
\ee
with the dots representing terms of higher order in $\d_\mu$. The function $\Gamma(\phi_0)$ is the effective potential to be used for the study of spontaneous symmetry breaking beyond the classical approximation. Note there are no one loop corrections proportional to $\left(\d_\mu \phi_0\right)^2$, making $Z(\phi_0)=1$ in this approximation.

%%%%%%%%%%%%%%%%%%%%%%%%%%%%%%%%%%%
\begin{figure}[!t]
\begin{center}
\begin{tikzpicture}
[line width=1.5 pt, scale=0.8]

\begin{scope}
[shift={(-9.5,-0.15)}]
\node at (-0.5,0.15) {$ \Gamma=$};
\end{scope}

\begin{scope}
[shift={(-9.5,-1.5)}]
\node at (-0,0.15) {$=$};
\end{scope}

\begin{scope}
[shift={(-9.9,-2.25)}]
\node at (-0,0.15) {$^{\bf REG}=$};
\end{scope}

\begin{scope}
[shift={(-9.9,-3)}]
\node at (-0,0.15) {$^{\bf REN}=$};
\end{scope}

\begin{scope}
[shift={(-8.8,-0.15)}]
\draw [fill=black](0,0.15) circle (0.05cm);
\node at (0.65,0.15) {$ +$};
\end{scope}

\begin{scope}
[shift={(-8.8,-1.5)}]
\node at (0,0.15) {$\Lambda$};
\node at (0.65,0.15) {$ +$};
\end{scope}

\begin{scope}
[shift={(-8.8,-2.3)}]
\node at (0,0.15) {$\Lambda$};
\node at (0.65,0.15) {$ +$};
\end{scope}

\begin{scope}
[shift={(-8.8,-3.05)}]
\node at (0,0.15) {$\Lambda$};
\node at (0.65,0.15) {$ +$};
\end{scope}

\begin{scope}
[shift={(-7.8,-0.15)}]
\draw [fill=black] (0,0.15)--(1,0.15);
\node at (1.4,0.15) {$ +$};
\end{scope}

\begin{scope}
[shift={(-7.8,-1.5)}]
\node at (0.5,0.15) {$ \frac{m^2}{2}\phi_0^2$};
\node at (1.4,0.15) {$ +$};
\end{scope}

\begin{scope}
[shift={(-7.8,-2.3)}]
\node at (0.5,0.15) {$ \frac{m^2}{2}\phi_0^2$};
\node at (1.4,0.15) {$ +$};
\end{scope}

\begin{scope}
[shift={(-7.8,-3.05)}]
\node at (0.5,0.15) {$ \frac{m^2}{2}\phi_0^2$};
\node at (1.4,0.15) {$ +$};
\end{scope}

\begin{scope}
[shift={(-5.8,0.05)}]
\draw [fill=black] (-0.3,0.3)--(0.3,-0.3);
\draw [fill=black] (0.3,0.3)--(-0.3,-0.3);
\draw [fill=black](0,0) circle (0.05cm);
\node at (0.65,0) {$ +$};
\end{scope}

\begin{scope}
[shift={(-5.8,-1.5)}]
\node at (0,0.15) {$ \frac{\lambda}{4!}\phi_0^4$};
\node at (0.75,0.15) {$ +$};
\end{scope}

\begin{scope}
[shift={(-5.8,-2.3)}]
\node at (0,0.15) {$ \frac{\lambda}{4!}\phi_0^4$};
\node at (0.75,0.15) {$ +$};
\end{scope}

\begin{scope}
[shift={(-5.8,-3.05)}]
\node at (0,0.15) {$ \frac{\lambda}{4!}\phi_0^4$};
\node at (0.75,0.15) {$ +$};
\end{scope}

\begin{scope}
[shift={(-4.25,-0.15)}]
\draw (0.5,0.15) circle (0.25cm);
\node at (1.6,0.15) {$ +$};
\end{scope}

\begin{scope}
[shift={(-4,-1.5)}]
\node[font=\tiny] at (0.25,0.15) {$ \frac{i}{2} \tilde{\int}\ln{\left[\frac{1}{l^2+m^2}\right]}$};
\node at (1.45,0.15) {$ -$};
\end{scope}

\begin{scope}
[shift={(-4.65,-2.3)}]
\node[font=\tiny] at (0,0.15) {$X$};
\node at (0.4,0.15) {$ -$};
\end{scope}

\begin{scope}
[shift={(-4.65,-3.05)}]
\node at (1,0.15) {$0$};
\node at (2,0.15) {$ +$};
\end{scope}

\begin{scope}
[shift={(-2,-0.15)}]
\draw [fill=black] (0,0)--(1,0);
\draw [fill=black](0.5,0) circle (0.05cm);
\draw (0.5,0.25) circle (0.25cm);
\end{scope}

\begin{scope}
[shift={(-0.3,0)}]
\node at (0,0) {$ +$};
\end{scope}

\begin{scope}
[shift={(-2,-1.5)}]
\node[font=\tiny]at (0.5,0.15) {$ \frac{i\lambda \phi_0^2}{4}\tilde{\int} \frac{1}{l^2+m^2}$};
\end{scope}

\begin{scope}
[shift={(-0.3,-1.35)}]
\node at (0,0) {$ +$};
\end{scope}

\begin{scope}
[shift={(-2.75,-2.3)}]
\node[font=\tiny] at (0.5,0.15) {$\frac{\lambda \phi_0^2 m^2}{64\pi^2} \left[ \frac{1}{\eps}+1+\ln{\frac{\mu^2}{m^2}}  \right] $};
\end{scope}

\begin{scope}
[shift={(-2.75,-3.05)}]
\node at (1.25,0.15) {$0$};
\end{scope}

\begin{scope}
[shift={(-0.3,-2.15)}]
\node at (0,0) {$ -$};
\end{scope}

\begin{scope}
[shift={(-0.3,-2.9)}]
\node at (0,0) {$ +$};
\end{scope}

\begin{scope}
[shift={(0.8,0)}]
\draw (0,0.25)--(0.25,0);
\draw (0,-0.25)--(0.25,0);
\draw [fill=black](0.25,0) circle (0.05cm);
\draw (0.5,0) circle (0.25cm);
\draw [fill=black](0.75,0) circle (0.05cm);
\draw (0.75,0)--(1,0.25);
\draw (0.75,0)--(1,-0.25);
\end{scope}

\begin{scope}
[shift={(3.1,0)}]
\node at (0,0) {$ +$};
\end{scope}

\begin{scope}
[shift={(0.8,-1.6)}]
\node[font=\tiny] at (0.5,0.15) {$ \frac{i\lambda^2 \phi_0^4}{16}\tilde{\int} \frac{1}{\left(l^2+m^2\right)^2}$};
\end{scope}

\begin{scope}
[shift={(3.1,-1.35)}]
\node at (0,0) {$ -$};
\end{scope}

\begin{scope}
[shift={(1.1,-2.3)}]
\node[font=\tiny] at (0.35,0.15) {$\frac{\lambda^2\phi_0^4}{2^8\pi^2} \left[ \frac{1}{\eps}+\ln{\frac{\mu^2}{m^2}}  \right] $};
\end{scope}

\begin{scope}
[shift={(3.1,-2.15)}]
\node at (0,0) {$ +$};
\end{scope}

\begin{scope}
[shift={(1.1,-3.05)}]
\node at (0.35,0.15) {$0$};
\end{scope}

\begin{scope}
[shift={(3.1,-2.9)}]
\node at (0,0) {$ +$};
\end{scope}

\begin{scope}
[shift={(4.05,0)}]
\draw (0,0.25)--(0.25,0);
\draw (0,-0.25)--(0.25,0);
\draw [fill=black](0.25,0) circle (0.05cm);
\draw (0.5,0) circle (0.25cm);
\draw [fill=black](0.75,0) circle (0.05cm);
\draw (0.75,0)--(1,0.25);
\draw [fill=black](0.5,0.25) circle (0.05cm);
\draw (0.5,0.25)--(0.75,0.5);
\draw (0.5,0.25)--(0.25,0.5);
\draw (0.75,0)--(1,-0.25);
\node at (2.1,0) {$+$};
\end{scope}

\begin{scope}
[shift={(4.15,-1.6)}]
\node[font=\tiny] at (0.5,0.15) {$ \frac{i\lambda^3 \phi_0^6}{48}\tilde{\int} \frac{1}{\left(l^2+m^2\right)^3}$};
\node at (2,0.25) {$ +$};
\end{scope}

\begin{scope}
[shift={(4.35,-2.3)}]
\node[font=\tiny] at (0.5,0.15) {$\frac{\lambda^3\phi_0^6}{3\cdot 2^9\pi^2m^2} $};
\node at (1.8,0.15) {$ -$};
\end{scope}

\begin{scope}
[shift={(4.35,-3.05)}]
\node[font=\tiny] at (0.5,0.15) {$\frac{\lambda^3\phi_0^6}{3\cdot 2^9\pi^2m^2} $};
\node at (1.8,0.15) {$ -$};
\end{scope}

\begin{scope}
[shift={(7.2,0)}]
\draw (0,0.25)--(0.25,0);
\draw (0,-0.25)--(0.25,0);
\draw [fill=black](0.25,0) circle (0.05cm);
\draw (0.5,0) circle (0.25cm);
\draw [fill=black](0.75,0) circle (0.05cm);
\draw (0.75,0)--(1,0.25);
\draw [fill=black](0.5,0.25) circle (0.05cm);
\draw [fill=black](0.5,-0.25) circle (0.05cm);
\draw (0.5,0.25)--(0.75,0.5);
\draw (0.5,0.25)--(0.25,0.5);
\draw (0.5,-0.25)--(0.75,-0.5);
\draw (0.5,-0.25)--(0.25,-0.5);
\draw (0.75,0)--(1,-0.25);
\node at (2.4,0) {$ +~\dots$};
\end{scope}

\begin{scope}
[shift={(7.2,-1.6)}]
\node[font=\tiny] at (0.5,0.15) {$ \frac{i\lambda^4 \phi_0^8}{128}\tilde{\int} \frac{1}{\left(l^2+m^2\right)^4}$};
\node at (2.4,0.25) {$ +~\dots$};
\end{scope}

\begin{scope}
[shift={(7.2,-2.3)}]
\node[font=\tiny] at (0.5,0.15) {$\frac{\lambda^4\phi_0^8}{3\cdot 2^{12}\pi^2m^4} $};
\node at (2.4,0.25) {$ +~\dots$};
\end{scope}

\begin{scope}
[shift={(7.2,-3.05)}]
\node[font=\tiny] at (0.5,0.15) {$\frac{\lambda^4\phi_0^8}{3\cdot 2^{12}\pi^2m^4} $};
\node at (2.4,0.25) {$ +~\dots$};
\end{scope}

\begin{scope}
[shift={(7.2,-3.8)}]
\node at (-0.6,0.15) {$\underbrace{\qquad\qquad\qquad\qquad\qquad\qquad\qquad\qquad}$};
\node[font=\tiny] at (-0.25,-0.6) {$\frac{1}{64\pi^2}\left(  \left(m^2+\lambda\phi_0^2/2\right)^2\ln{\left[ 1+\frac{\lambda\phi_0^2/2}{m^2} \right]}
   -\frac{\lambda\phi_0^2}{8}\left(4m^2+3\lambda\phi_0^2\right)\right)$};
\end{scope}

\end{tikzpicture}
\end{center}

\caption{The tree and one-loop diagrams that contribute to the effective action. $\tilde{\int} \equiv \int\frac{d^4l}{(2\pi)^4}$. ``REG" denotes (dimensional) regularisation, ``REN" denotes renormalisation (in this case, following the conditions in eq.~\ref{ren4v}.) $x$ denotes the irrelevant (independent of $\phi_0$) contribution to the zero-point function coming from $ \frac{i}{2} \tilde{\int}\ln{\left[\frac{1}{l^2+m^2}\right]}$. The Planck constant $\hbar$ is set to one.}
\label{gamdiag}
\end{figure}

%%%%%%%%%%%%%%%%%%%%%%%%%%%%%%

The tree and one-loop diagrams for the effective potential are shown in Fig.~\ref{gamdiag}. For two and four external legs the one-loop contributions are divergent (and all further loop corrections as well) and can be regulated in several ways. We will use dimensional regularisation for definiteness.

Next, one chooses a renormalisation scheme to instruct how these divergences should be removed. As always, different renormalisation schemes will lead to different results for the effective potential, but the resulting physical predictions are equal, at least up to next order in perturbation theory. Our choice is to impose that at a given renormalisation point, all quantum corrections to the effective potential vanish\footnote{Yet another common choice is to define the condition on $d^4 \Gamma / d \phi_0^4$ at an arbitrary ``renormalisation scale" $\mu$. This approach connects closely to the renormalisation group. For our purposes we find the scheme in eqs.~\ref{ren4v} more instructive.}. To that end, we need to demand
\ba
\frac{d^2 \Gamma}{d \phi_0^2} \Big|_{\phi_0=\phi_0^\star} = m^2+\frac{\lambda \phi_0^{\star 2}}{2}\,,~~~~~
 \frac{d^4 \Gamma}{d \phi_0^4} \Big|_{\phi_0=\phi_0^\star} = \lambda,
\label{ren4v}
\ea
where $\Gamma$ denotes the renormalised effective potential, and $\phi_0^\star$ denotes the renormalisation point: the value for $\phi_0$ at which the renormalisation conditions are imposed. Note that since we want quantum corrections to vanish at the renormalisation point $\phi_0^\star$, the right hand sides of the conditions in eqs.~\ref{ren4v} are simply obtained from taking two or four derivatives of the classical potential $\frac{m^2 \phi_0^2}{2} +\frac{\lambda\phi_0^4}{4!}$.

To execute this renormalisation program, one introduces counter-terms $\delta Z$, $\delta \Lambda$ $\delta m$ and $\delta \lambda$, and finds their corresponding values that lead to the renormalised effective action that meets the conditions above. At any rate, one ends up with
\ba
i\cdot \Gamma_{\rm eff} &=&  \int d^4 x \Biggl[\frac{1}{2}\left(\d_\mu \phi_0\right)^2 +\frac{m^2}{2}\phi_0^2 +\frac{\lambda}{4!}\phi_0^4 \nn\\
&& \qquad  \qquad+\hbar\cdot\frac{m^4}{32\pi^2}\cdot \sum_{n=3}^\infty \left(-1\right)^{n+1}\cdot \frac{1}{n\cdot (n-1)\cdot(n-2)}\cdot\left(\frac{\lambda \phi_0^2}{2m^2}\right)^n +\dots \Biggr]
\ea 
and ultimately\footnote{One could add an extra term $\alpha_0\cdot m^4$ to cancel a constant, $\phi_0$-independent contribution from the first term, but until Appendix \ref{cc} we will not bother about unphysical constant contributions to the effective action.}
\ba
i\cdot \Gamma_{\rm eff} &=&   \int d^4 x \Biggl[\frac{1}{2}\left(\d_\mu \phi_0\right)^2 +\frac{m^2}{2}\phi_0^2 +\frac{\lambda}{4!}\phi_0^4\nn\\
&& \qquad \qquad +\frac{\hbar}{64\pi^2}\left(\left(m^2+\lambda\phi_0^2/2\right)^2\ln{\left[ 1+\frac{\lambda\phi_0^2/2}{m^2} \right]} - \alpha_1 \cdot m^2 \phi_0^2 - \alpha_2 \cdot \phi_0^4
\right)\Biggr], \label{ve}
\ea
with the coefficients $\alpha_1$ and $\alpha_2$ to be fixed by the renormalisation procedure (more precisely, by the choice for the classical field value $\phi_0^\star$ at which the renormalisation conditions in eqs.~\ref{ren4v} are imposed).

\subsection{Symmetric phase}
\label{symm}

In the symmetric phase ($m^2>0$) a common choice is to impose the renormalisation conditions at $\phi_0=0$:
\ba
\frac{d^2 \Gamma}{d \phi_0^2} \Big|_{\phi_0=0} = m^2\,,~~
 \frac{d^4 \Gamma}{d \phi_0^4} \Big|_{\phi_0=0} = \lambda. \label{een}
\ea
In this case, one finds that the coefficients $\alpha_i$ in the general result of eq.~\ref{ve} are given by
\be
\alpha_1= \frac{\lambda}{2}\,, \qquad \alpha_2 = \frac{3\lambda^2}{8}.\label{vecoeff}
\ee
%Note that in this particular case, the solution above corresponds to simply throwing away the one-loop corrections for $n=2$ and $n=4$ entirely. 

\subsection{Broken phase}
\label{brok}
When $m^2 <0$, spontaneous symmetry breaking occurs and the classical potential finds a minimum at a nonzero field value
\be
\phi_0 = \pm \sqrt{-\frac{6m^2}{\lambda}}   \equiv \pm v\,.
 \label{singlevev}
\ee
In this situation, one may decide to compute the effective potential around this classical vacuum expectation value $\phi_0=v$. Now the conditions in eqs.~\ref{ren4v} read
\ba
\frac{d^2 \Gamma}{d \phi_0^2} \Big|_{\phi_0=v} = -2m^2\,,~~
\frac{d^4 \Gamma}{d \phi_0^4} \Big|_{\phi_0=v} = \lambda\,.
\label{ren4vssb}
\ea
As noted before, the expressions at the right-hand side are simply given by the corresponding number of derivatives acting on the classical potential, evaluated at the vev $v$. %In other words, the only fundamental difference with the previous section is that now we impose that quantum corrections vanish at $\phi_0=v$ rather than at $\phi_0=0$. 
In this case the coefficients $\alpha_1$ and $\alpha_2$ are given by
\be
%\alpha_1=\frac{285}{2}+i\pi+\ln{2}  \qquad \qquad
\alpha_1 = \lambda\left(\frac{61}{2}+i\pi+\ln{2}\right)\,,  \qquad \alpha_2 = \frac{\lambda^2}{4}\left(\frac{9}{2}+i\pi+\ln{2}\right)\,. \label{brocof}
\ee
Here the factors of $i\pi$ indicate that in a region $\phi_0^2 <-2m^2/\lambda$ around $\phi_0=0$ the effective potential develops an imaginary part. At $\phi_0^2>-2m^2/\lambda$ the imaginary part disappears.

Alternatively, one may replace the first condition in eqs.~\ref{ren4vssb} by a condition on the first derivative of the effective action:\footnote{Note that in the symmetric case, this condition makes no sense. Since the first derivative vanishes identically at $\phi_0=0$, the condition is void. Consequently, we are one condition short to renormalise the theory.}
\be
\frac{d \Gamma}{d \phi_0} \Big|_{\phi_0=v} = 0\,. 
\label{rem}
\ee
This condition ensures that quantum corrections do not push away the vev from its classical location in field space.  Again, we obtain the general result of eq.~\ref{ve}, this time with coefficients
\be
\alpha_1= \lambda \left(\frac{25}{2}+i\pi+\ln{2}\right) \qquad \qquad \alpha_2 = \frac{\lambda^2}{4}\left(\frac{9}{2}+i\pi+\ln{2}\right).
\label{brocof2}
\ee

\section{Finite approach} 
\label{fin}
 {The goal of this Section is to show that there exists an alternative way to compute the effective potential. Contrary to the traditional approach that we have just reviewed, this method never runs into any intermediate divergent quantity. In our opinion, the existence of such a method is of great conceptual interest. It counters the usual intuition that infinities are a fact of life in particle physics.}

\subsection{Setup}
 {This first Subsection introduces the finite CS method that this paper employs. In Subsubsection \ref{corrf}, it is introduced in the original context: in terms of correlation functions. That was the formulation that we used in our papers \cite{Mooij:2021ojy, Mooij:2021lbc}. Next, in Subsubsection \ref{potfin}, we formulate the method in terms of the effective potential. In \ref{effcs} we give the equations that we will use later on to compute the effective potential in a manifestly finite way.}

\subsubsection{Correlation functions} \label{corrf}

The finite QFT method to compute $n$-point correlation functions was presented in lecture notes for the Les Houches summer school in 1975 by Callan \cite{Callan:1975vs}. It was applied to quantum electrodynamics in \cite{Blaer:1974foy}. Its central equations are similar (but not equal to) the Callan-Symanzik equation \cite{Callan:1970yg, Symanzik:1970rt}. Therefore we refer to it as the ``Callan-Symanzik (CS) method". It rests on a so-called ``$\theta$-operation". Graphically it can be thought of as cutting a propagator in two and pasting them together by a ``$\theta$-vertex", which comes with Feynman rule $(-1)$. Algebraically, it acts on an $n$-point correlation function through
\ba
\tilde{\Gamma}^{(n)}_\theta \equiv -i \times \frac{d}{d m_0^2} \tilde{\Gamma}^{(n)}\,,\qquad \tilde{\Gamma}^{(n)}_{\theta\theta} \equiv -i \times \frac{d}{d m_0^2} \tilde{\Gamma}_\theta^{(n)}\,.
\label{thetdef}
\ea
Here $\tilde{\Gamma}$ denotes the bare Green's functions { , while $m_0$ denotes the bare mass. Bare quantities} show up only in the derivation of the CS method, not in its actual use\footnote{ {As we are only concerned with executing the CS program rather than deriving it, there is no need to consider bare quantities at all. The point of the CS operation is precisely that once established, it circumvents the entire notion of bare quantities.}}. The $\theta$-operation lowers the degree of divergence by two units. The finite correlation functions $\Gamma^{(n)}$, $\Gamma^{(n)}_\theta$ and $\Gamma^{(n)}_{\theta\theta}$ are connected through \cite{Callan:1975vs}
\ba
2i m^2 ~(1+\gamma)~\Gamma^{(n)}_\theta &=&  \left(2m^2 ~\frac{\d}{\d m^2} +\beta~\frac{\d}{\d \lambda}    +n\cdot \gamma \right) \Gamma^{(n)}~,\nn\\
2im^2~(1+\gamma)\cdot  \Gamma^{(n)}_{\theta\theta} &=& \left(2m^2~\frac{\d}{\d m^2} +\beta~\frac{\d}{\d \lambda} +n\cdot \gamma +\gamma_\theta\right)\Gamma^{(n)}_\theta. \label{cwold}
\ea
Here $\beta$, $\gamma$ and $\gamma_\theta$ are additional parameters that find their values in the course of computing the correlation functions\footnote{ {Note that a ``fermionic version" of the CS approach would involve cutting fermion propagators in two by taking derivatives with respect to one power of the fermion mass instead of two. Such an operation would lower the degree of divergence by one unit \cite{Blaer:1974foy}.}}.  {$\beta$ and $\gamma$ are very reminiscent of the usual beta-functions and anomalous dimension, but we stress that in this context they are just parameters. The same goes for the CS equations themselves: they are reminiscent of the famous Callan-Symanzik equation used to compute the high energy behaviour of the Green's functions, but different. In partucular, they do not contain the derivatives with respect to momenta. Furthermore, the parameter ``$n$" is the same as the ``$n$" appearing in the superscripts of the correlation functions.}
%%%%%%%%%%%%%%%%%%%%%%%%%%%%%
\begin{figure}[!b]
 \centering
    \includegraphics[width=0.9\textwidth]{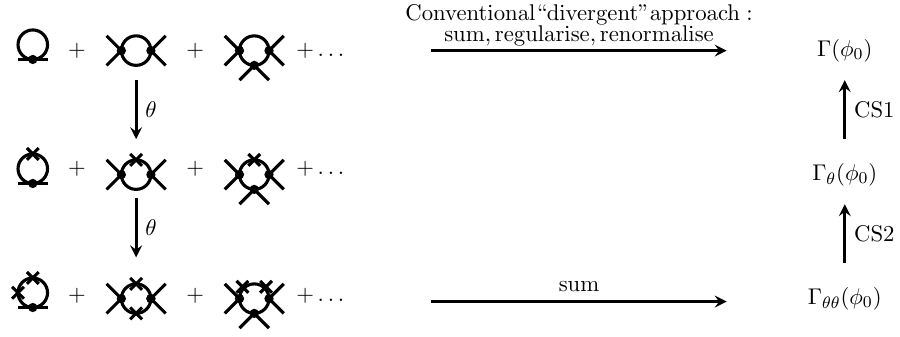}
\caption{Schematic depiction of the finite derivation of the one-loop corrections to the effective action. CS1 and CS2 represent the first and second CS equations in eqs.~\ref{effcs}.}
\label{finfig}
\end{figure}
%%%%%%%%%%%%%%%%%%%%%%%%%%%%%%%

Of course, one could continue and define correlation functions with any number of $\theta$-subscripts, and the corresponding equations. However, in the $\lambda \phi^4$ theory, two $\theta$-operations are enough to render all $n$-point correlation functions with $n>0$ convergent. The set of equations in eqs.~\ref{cwold} then allows for a computation, up to any desired loop order, of the remaining correlation functions with positive or zero formal degree of divergence (i.e., $\Gamma^{(2)}$, $\Gamma^{(4)}$ and $\Gamma^{(2)}_\theta$) {\it without ever running into a divergent quantity}. Therefore, at no point it is required to perform ``divergent subtractions" to render the correlation functions finite.

\subsubsection{Effective potential} \label{potfin}

From eq.~\ref{pos} it is direct to see that the set of equations in \ref{cwold} can be straightforwardly generalised from $n$-point correlation functions $\Gamma^{(n)}$ to the effective potential $ \Gamma(\phi_0)$ \cite{Mooij:2021ojy}:
\ba
2 im^2 ~(1+\gamma) ~ \Gamma_{\theta}&=&  \left(2m^2 ~\frac{\d}{\d m^2} +\beta~\frac{\d}{\d \lambda}    + \gamma \cdot \phi_0\frac{\d}{\d \phi_0}\right)  \Gamma,\nn\\
2im^2~(1+\gamma) ~   \Gamma_{\theta\theta} &=& \left(2m^2~\frac{\d}{\d m^2} +\beta~\frac{\d}{\d \lambda} +\gamma \cdot \phi_0\frac{\d}{\d \phi_0}+\gamma_\theta\right) \Gamma_{\theta}, \label{effcs}
\ea
where we have rewritten the factor of $n$ in the first set of equations through a field derivative. In terms of bare quantities, one can define $\Gamma_\theta$ and $\Gamma_{\theta\theta}$ following eq.~\ref{thetdef} 
\ba
\tilde{\Gamma}_\theta \equiv -i \times \frac{d}{d m_0^2} \tilde{\Gamma}\,,\qquad \tilde{\Gamma}_{\theta\theta} \equiv -i \times \frac{d}{d m_0^2} \tilde{\Gamma}_\theta\,.
\label{thetdef2}
\ea

As mentioned above, even if the $\theta$-operation is defined in eqs.~\ref{thetdef} on bare correlation functions, all computations rely on the system of two equations above, which relate only the finite quantities. 

Note that the equations above are in terms of the effective potential $\Gamma(\phi_0)$, rather than in terms of the effective action $\Gamma_{\rm eff}$. Of course, the CS equations can be rewritten in terms of (functional derivatives of) the effective action, making our discussion more complete.
%However, at one loop level, there is no field renormalisation in $\lambda \phi^4$ theory. The effective action (and its $\theta$-descendents) are then related through the effective potential (and its $\theta$-descendents) through
%\ba
%\Gamma_{\rm eff} &=& -i\int d^4 x~ \Gamma(\phi_0)\nn\\
%\Gamma_{{\rm eff},\theta} &=& -i\int d^4 x~ \Gamma_\theta(\phi_0)\nn\\
%\Gamma_{{\rm eff},\theta\theta} &=& -i\int d^4 x ~ %\Gamma_{\theta\theta}(\phi_0). \label{gamsimp}
%\ea
%Of course, without switching from $\{\Gamma_{\rm eff},\Gamma_{{\rm eff},\theta}, \Gamma_{{\rm eff},\theta\theta}\}$ to $\{\Gamma,\Gamma_\theta,\Gamma_{\theta\theta}\}$ our discussion would be more complete. 
%When moving to second loop order and beyond, wavefield renormalisation appears and the discussion should really be presented in terms of the full effective action. 
In spirit, nothing changes in the CS framework, but computations (and notations) become much more intricated. For the purposes of our paper (in particular, for the proof of principle that we want to provide), we prefer to present the simplified discussion in terms of just $\{\Gamma,\Gamma_\theta,\Gamma_{\theta\theta}\}$.

Armed with the standard boundary conditions of eqs.~\ref{ren4v}, we are now ready to derive the effective potential. We omit the derivation (as it is identical to the one presented in \cite{Callan:1975vs}) that shows that $\beta$ must begin at order $\lambda^2$, while the $\gamma$ and $\gamma_\theta$ can have support at order $\lambda$ as well.

\subsection{Computation}

 {We now get to the most technical part of the paper. We want to show explicitly that the equations derived in \ref{effcs} allow for a manifestly finite derivation of the effective potential. The endpoint of the computation is in eq.~\ref{endpoint}. From there on we are back on the ``traditional path" (i.e., the standard approach described in Section \ref{stapp}) towards the standard result for the effective potential, eq.~\ref{ve}. Again, the central arguments of this paper are of conceptual nature. The physics ideas that we want to convey do not hinge on the details of the computation that we are about to present. In spirit, this Subsection can be summarised as just ``The effective potential can indeed be derived by the finite CS method." }

We begin from defining
\be
\Gamma = \Gamma_0+\hbar\cdot \phi_0^4 \cdot \Gamma_1\left(\phi_0^2/m^2\right)+\Ohh.
\ee
Here $\Gamma_0$ denotes the classical potential
\be
\Gamma_0 = \frac{m^2 \phi_0^2}{2} + \frac{\lambda \phi_0^4}{4!},
\ee
while the dimensionless function $\Gamma_1$ is the one-loop correction to the effective potential that we are after.

Similarly, we define
\ba
\Gamma_\theta &=& \Gamma_{\theta,0} +\hbar\cdot \Gamma_{\theta,1}+\Ohh\,,\nn\\
\Gamma_{\theta\theta} &=& \Gamma_{\theta\theta,0} +\hbar\cdot \Gamma_{\theta \theta,1}+\Ohh\,,
\ea
and also
\be
\beta = \hbar\cdot \beta_1 + \Ohh, \qquad \qquad \gamma = \hbar\cdot \gamma_1 + \Ohh,\qquad \qquad \gamma_\theta = \hbar\cdot \gamma_{\theta,1} + \Ohh.
\ee
Evaluating the first CS equation in \ref{effcs} at tree level order (i.e., at zeroth order in $\hbar$) gives
\be
\Gamma_{\theta,0} = -i\cdot\frac{\phi_0^2}{2}\,.\label{gtnul}
\ee
With that, we can evaluate the second CS equation in \ref{effcs} at tree level in order to find
\be
\Gamma_{\theta\theta,0} = 0\,.
\ee
Actually, now that we have decided to leave out the vacuum bubble diagram, all contributions to $\Gamma_{\theta \theta}$ are finite. Therefore, it can be computed without encountering intermediate divergences. See Fig.~\ref{gttfig}. This observation forms the core of the CS mechanism. At one loop (i.e., order $\hbar$) we get
%\ba
%\Gamma_{\theta\theta} &=& \frac{i\hbar}{32\pi^2}\left(\frac{\lambda \phi_0^2}{2m^2} -\frac{\lambda^2 \phi_0^4}{8 m^4} +\frac{\lambda^3\phi_0^6}{24 m^6}+\dots \right)= \frac{i\hbar}{32\pi^2}\times \ln{\left[1+\frac{\lambda \phi_0^2}{2m^2} \right]}.
%
%\ea
%%%%%%%%%%%%%%%%%%%%%%%%%%%%%%%%%%%%%
%\begin{figure}[!b]
 %\centering
  %  \includegraphics[width=0.8\textwidth]{Fig3}
%\caption{One loop contributions to $\Gamma_{\theta\theta}$. Classical fields $\phi_0$ on the external lines, quantum fields $h$ on the internal lines. $\theta$-vertex (with Feynman rule $-1$) denoted by little crosses. $\tilde{\int} \equiv \int\frac{d^4l}{(2\pi)^4}$. $\hbar$ set to one. Vacuum diagram omitted.}
%\label{gttfig}
%\end{figure}
%%%%%%%%%%%%%%%%%%%%%%%%%%%%%

\begin{figure}[!t]
\begin{center}
\begin{tikzpicture}
[line width=1.5 pt, scale=1.5]

\begin{scope}
[shift={(-3.85,-0.15)}]
\node at (0,0.15) {$ \Gamma_{\theta\theta,1}$};
\end{scope}

\begin{scope}
[shift={(-3.25,0)}]
\node at (0,0) {$=$};
\node at (0,-0.75) {$=$};
\node at (0,-1.5) {$=$};
\node at (0,-2.25) {$=$};
\end{scope}

\begin{scope}
[shift={(-2,0)}]
\node at (-0.5,0) {$2~\times$}; 
\draw [fill=black] (0,0)--(1,0);
\draw [fill=black](0.5,0) circle (0.05cm);
\draw (0.5,0.25) circle (0.25cm);
\draw [fill=black] (0.4,0.6)--(0.6,0.4);
\draw [fill=black] (0.4,0.4)--(0.6,0.6);
\draw [fill=black] (0.15,0.35)--(0.35,0.15);
\draw [fill=black] (0.15,0.15)--(0.35,0.35);
%\node at (1.65,0.15) {$ +$};
%\node at (2.2,0.15) {$ 6~~\times$};
\end{scope}

\begin{scope}
[shift={(-1.7,-0.75)}]
\node at (0,0) {$\frac{i\lambda \phi_0^2}{2}\tilde{\int} \frac{1}{\left(l^2+m^2\right)^3}$}; 
\end{scope}

\begin{scope}
[shift={(-1.7,-1.5)}]
%\node at (-0.25,0){-}
\node at (0,0) {$-\frac{\lambda \phi_0^2}{64\pi^2\cdot m^2}$};  
\end{scope}

\begin{scope}
[shift={(-1.7,-2.25)}]
\node at (0,0) {$-\frac{1}{32\pi^2}\cdot \ln{\left[1+\frac{\lambda \phi_0^2}{2m^2}\right]}$};  
\end{scope}

\begin{scope}
[shift={(-0.5,0)}]
\node at (0,0) {$+$}; 
\node at (0,-0.75) {$-$}; 
\node at (0,-1.5) {$+$}; 
\end{scope}

\begin{scope}
[shift={(0.5,0)}]
\node at (-0.5,0) {$6~\times$}; 
\draw (0,0.25)--(0.25,0);
\draw (0,-0.25)--(0.25,0);
\draw [fill=black](0.25,0) circle (0.05cm);
\draw (0.5,0) circle (0.25cm);
\draw [fill=black] (0.4,0.35)--(0.6,0.15);
\draw [fill=black] (0.4,0.15)--(0.6,0.35);
\draw [fill=black] (0.4,-0.15)--(0.6,-0.35);
\draw [fill=black] (0.4,-0.35)--(0.6,-0.15);
\draw [fill=black](0.75,0) circle (0.05cm);
\draw (0.75,0)--(1,0.25);
\draw (0.75,0)--(1,-0.25);
%\node at (1.65,0) {$ +$};
\node at (2.2,0) {$ 12~~\times$};
\end{scope}

\begin{scope}
[shift={(0.8,-0.75)}]
\node at (0,0) {$\frac{3i\lambda^2\phi_0^4}{8} \tilde{\int} \frac{1}{\left(l^2+m^2\right)^4}$}; 
\end{scope}

\begin{scope}
[shift={(0.8,-1.5)}]
\node at (0,0) {$\frac{\lambda^2\phi_0^4}{2^8\pi^2\cdot m^4}$};  
\end{scope}

\begin{scope}
[shift={(2.15,0)}]
\node at (0,0) {$+$}; 
\node at (0,-0.75) {$+$}; 
\node at (0,-1.5) {$-$}; 
\end{scope}

\begin{scope}
[shift={(3,0)}]
\draw (0,0.25)--(0.25,0);
\draw (0,-0.25)--(0.25,0);
\draw [fill=black](0.25,0) circle (0.05cm);
\draw (0.5,0) circle (0.25cm);

\draw [fill=black] (0.33-0.1,0.17+0.1)--(0.33+0.1,0.17-0.1);
\draw [fill=black] (0.33+0.1,0.17+0.1)--(0.33-0.1,0.17-0.1);

\draw [fill=black] (0.4,-0.15)--(0.6,-0.35);
\draw [fill=black] (0.4,-0.35)--(0.6,-0.15);
\draw [fill=black](0.75,0) circle (0.05cm);
\draw (0.75,0)--(1,0.25);
\draw [fill=black](0.5,0.25) circle (0.05cm);
\draw (0.5,0.25)--(0.75,0.5);
\draw (0.5,0.25)--(0.25,0.5);
\draw (0.75,0)--(1,-0.25);
\node at (1.65,0) {$ +~~~~~\dots$};
\end{scope}

\begin{scope}
[shift={(3,-0.75)}]
\node at (0.25,0) {$\frac{i\lambda^3 \phi_0^6}{4}\tilde{\int} \frac{1}{\left(l^2+m^2\right)^5}$};
\node at (1.65,0) {$ +~~~~~\dots$};
\end{scope}

\begin{scope}
[shift={(3,-1.5)}]
\node at (0.25,0) {$\frac{\lambda^3 \phi_0^6}{3\cdot 2^8\pi^2 m^6}$};
\node at (1.65,0) {$ +~~~~~\dots$};
\end{scope}

\end{tikzpicture}
\end{center}

\caption{One loop contributions to $\Gamma_{\theta\theta,1}$. Classical fields $\phi_0$ on the external lines, quantum fields $h$ on the internal lines. $\theta$-vertex (with Feynman rule $-1$) denoted by little crosses. $\tilde{\int} \equiv \int\frac{d^4l}{(2\pi)^4}$. $\hbar$ set to one. Vacuum diagram omitted.}
\label{gttfig}
\end{figure}
%%%%%%%%%%%%%%%%%%%%%%%%%%%%%%
\be
\Gamma_{\theta\theta,1} = -\frac{1}{32\pi^2} \log{\left[  1+\frac{\lambda \phi_0^2}{2m^2} \right]}\,.\label{gttb}
\ee

Next we evaluate the first CS equation in \ref{effcs} at order $\hbar$. In terms of a dimensionless variable $x$
\be
x\equiv \frac{\phi_0^2}{m^2}
\ee
we get
\be
\frac{\Gamma_{\theta,1}}{m^2} = i\cdot x^2 \left[x~\Gamma_1'(x)-\frac{\beta_1+4\gamma_1\lambda}{48}\right].
\ee
%That allows us to express $\Gamma_{\theta,1}$ in terms of $\Gamma'_1$, where the prime denotes differentiation with respect to the dimensionless variable $\phi_0^2/m^2$. We get
%\be
%2im^2\cdot \left(\Gamma_{\theta,1}-\frac{\gamma\phi_0^2}{2}\right) = -i\cdot \left(  2m^2 \Gamma_{\theta,1} + \frac{\beta+4\gamma\lambda}{4!}\phi_0^4  -\frac{2 \phi_0^6 \Gamma'_1}{m^2} \right).
%\ee
%\be
%\Gamma_{\theta,1} = \frac{i \phi_0^4}{m^4} \left[\phi_0^2 ~\Gamma_1'\left(\frac{\phi_0^2}{m^2}\right)-\frac{m^2\left(\beta_1+4\gamma_1\lambda\right)}{48}\right]
%\ee
The following step is to evaluate the second CS equation in eqs.~\ref{effcs} at first order in $\hbar$. Inserting the previous result, we end up with a second order differential equation for $\Gamma_1$:
\be
x^4 \Gamma_1''(x)+2x^3 \Gamma_1'(x) -\frac{x^2\left(\beta_1+4\gamma_1\lambda\right)}{48} +\frac{ x\left(2\gamma_1+\gamma_{\theta,1}\right)}{4}= \frac{\log{\left[1+\lambda x/2\right]}}{32\pi^2}.
\ee
We get
\ba
\Gamma_1 &=& \frac{1}{64\pi^2}\left[ \frac{\log{\left[1+\lambda x/2\right]}}{x^2} +\lambda~\log{\left[ 2+\lambda x\right]}\left(\frac{1}{x}+\frac{\lambda}{4}\right)  \right] \nn\\
&& \qquad \qquad +\log{[x]}\cdot \left[ \frac{1}{4x}\left(2\gamma_1+\gamma_{\theta,1}-\frac{\lambda}{16\pi^2}\right) +\frac{1}{48}\left(\beta_1+4\gamma_1\lambda-\frac{3\lambda^2}{16\pi^2}\right)   \right]\nn\\
&& \qquad \qquad +\frac{1}{4x}\left(2\gamma_1+\gamma_\theta-\frac{\lambda}{32\pi^2}+c_1\right) + c_2, \label{finres}
\ea
where $c_1$ and $c_2$ are two integration constants.

Now we use that at one loop, there is no field renormalisation, so we can safely set the order $\hbar$ contribution $\gamma_1\to 0$\footnote{Strictly speaking, at this point we do not know that this CS parameter $\gamma$ is just the usual field renormalisation parameter $\gamma$. However, we can take that fact from our paper \cite{Mooij:2021ojy} devoted to the CS computation of $n$-point functions $\Gamma^{(n)}$. With that, we understand that at one loop, $\gamma$ indeed disappears. See also the discussion below eq.~\ref{derexp}}. Next, we use that imposing analyticity (i.e. imposing that the solution is regular at $\phi_0=0$) forbids the terms proportional to $\log{(x)}$.  {The requirement of analyticity of the effective potential at small $\phi$ is nothing but the requirement of existence of Green's functions (Taylor coefficients in the expanding of the effective action) in the  $\lambda \phi^4$ theory with non-zero scalar mass. Though this is not guaranteed beyond perturbation theory (``triviality'' of the model related to the Landau pole in the scalar coupling constant), these Green's functions do exist perturbatively. }  With $\gamma$ already out of the way, we can then solve for the remaining two CS parameters:
\be
\beta_1 \to \frac{3\lambda^2}{16\pi^2},\qquad\qquad \gamma_{\theta,1} \to \frac{\lambda}{16\pi^2}. \label{bl}
\ee
With that, we are left with a well-behaving solution for $\Gamma_1$ (with newly defined integration constants $c_1$ and $c_2$):
\be
\phi_0^4 \cdot \Gamma_1 =\frac{1}{64\pi^2}\left[ \left(m^2+\frac{\lambda \phi_0^2}{2}\right)^2 \log{\left( 1+\frac{\lambda \phi_0^2}{2m^2}\right)}   \right] + c_1\cdot m^2\phi_0^2 + c_2\cdot \phi_0^4 . \label{endpoint}
\ee 
We have arrived on familiar grounds by now. By imposing two boundary conditions at a chosen field value we can solve for the integration constants $c_1$ and $c_2$. Everything works as in the standard computation: the boundary conditions of eq.~\ref{een} lead to the result in eqs.~\ref{ve} with the coefficients in eq.~\ref{vecoeff}, the boundary conditions of eq.~\ref{ren4vssb} lead to the same result in eq.~\ref{ve} but with the coefficients in eq.~\ref{brocof}, etcetera.

For completeness, we mention the result for $\Gamma_{\theta,1}$:
\be
\Gamma_{\theta,1} = -\frac{i}{32\pi^2}\left(m^2+\frac{\lambda\phi_0^2}{2}\right) \log{\left( 1+\frac{\lambda \phi_0^2}{2m^2}\right)} + c_1\cdot \phi_0^2,
\ee
where $c_1$ is an integration constant.\\
\\
As a ``divergent double check", one can of course compute $\Gamma_{\theta}$ through the conventional approach of regularising and renormalising divergent Feynman diagrams. This is illustrated in Fig.~\ref{gtfig} (for the particular case $\phis=0$).
%, using the renormalisation condition of footnote \ref{fn}).
%%%%%%%%%%%%%%%%%%%%%%%%%%
%\begin{figure}[!t]
 %\centering
  %  \includegraphics[width=0.8\textwidth]{Fig4}
%\caption{Double check: traditional ``divergent" computation of the tree level and one loop contributions to $\Gamma_\theta$. Classical fields $\phi_0$ on the external lines, quantum fields $h$ on the internal lines. $\theta$-vertex (with Feynman rule $-1$) denoted by little crosses. $\tilde{\int} \equiv \int\frac{d^4l}{(2\pi)^4}$. ``REG" denotes (dimensional) regularisation, ``REN" denotes renormalisation following the last condition in footnote \ref{fn} (for the case $\phis=0$). $\hbar$ set to one. Vacuum diagram omitted. }
%\label{gtfig}
%\end{figure}
%%%%%%%%%%%%%%%%%%%%%%%%%%%%%%%

\begin{figure}[!h]
\begin{center}
\begin{tikzpicture}
[line width=1.5 pt, scale=1.25]

\begin{scope}
[shift={(-6,0)}]
\node at (0,0) {$ \Gamma_\theta$};
\end{scope}

\begin{scope}
[shift={(-5.5,0)}]
\node at (0,0) {$=$};
\node at (0,-1) {$=$};
\node at (-0.25,-1.75) {$^{\bf {\rm REG}}=$};
\node at (-0.25,-2.5) {$^{\bf {\rm REN}}=$};
\end{scope}

\begin{scope}
[shift={(-5,0)}]
\draw [fill=black] (0,0)--(1,0);
\draw [fill=black] (0.4,0.1)--(0.6,-0.1);
\draw [fill=black] (0.4,-0.1)--(0.6,0.1);
\end{scope}

\begin{scope}
[shift={(-5,-1)}]
\node at (0.5,0) {$-\frac{i\phi_0^2}{2}$};
\node at (0.5,-0.75) {$-\frac{i\phi_0^2}{2}$};
\node at (0.5,-1.5) {$-\frac{i\phi_0^2}{2}$};
\end{scope}

\begin{scope}
[shift={(-3.5,0)}]
\node at (0,0) {$+$};
\node at (0,-1) {$+$};
\node at (0,-1.75) {$+$};
\node at (0,-2.5) {$+$};
\end{scope}

\begin{scope}
[shift={(-3,-0.15)}]
\draw [fill=black] (0,0)--(1,0);
\draw [fill=black](0.5,0) circle (0.05cm);
\draw (0.5,0.25) circle (0.25cm);
\draw [fill=black] (0.4,0.6)--(0.6,0.4);
\draw [fill=black] (0.4,0.4)--(0.6,0.6);
\end{scope}

\begin{scope}
[shift={(-2.35,-1)}]
\node at (0,0) {$\frac{\lambda \phi_0^2}{4}\tilde{\int}\frac{1}{\left(l^2+m^2\right)^2}$};
\end{scope}

\begin{scope}
[shift={(-2.45,-1.75)}]
\node[font=\tiny] at (0.2,0) {$\frac{i\lambda \phi_0^2}{64\pi^2} \left[\frac{1}{\eps}+\ln{\frac{\mu^2}{m^2}}\right]$};
\end{scope}

\begin{scope}
[shift={(-2.35,-2.5)}]
\node{$0$};
\end{scope}

\begin{scope}
[shift={(-0.75,0)}]
\node at (0,0) {$+~2~\times$};
\end{scope}

\begin{scope}
[shift={(-1,-1)}]
\node at (0,0) {$-$};
\node at (0,-0.75) {$-$};
\node at (0,-1.5) {$-$};
\end{scope}

\begin{scope}
[shift={(-0.25,0)}]
\draw (0,0.25)--(0.25,0);
\draw (0,-0.25)--(0.25,0);
\draw [fill=black](0.25,0) circle (0.05cm);
\draw (0.5,0) circle (0.25cm);
\draw [fill=black] (0.4,0.35)--(0.6,0.15);
\draw [fill=black] (0.4,0.15)--(0.6,0.35);
\draw [fill=black](0.75,0) circle (0.05cm);
\draw (0.75,0)--(1,0.25);
\draw (0.75,0)--(1,-0.25);
\end{scope}

\begin{scope}
[shift={(0.15,-1)}]
\node at (0,0) {$\frac{\lambda^2\phi_0^4}{8} \tilde{\int} \frac{1}{\left(l^2+m^2\right)^3}$};
\end{scope}

\begin{scope}
[shift={(0,-1.75)}]
\node at (0,0) {$\frac{i\lambda^2\phi_0^4}{2^8\pi^2 m^2}$};
\end{scope}

\begin{scope}
[shift={(0,-2.5)}]
\node at (0,0) {$\frac{i\lambda^2\phi_0^4}{2^8\pi^2 m^2}$};
\end{scope}

\begin{scope}
[shift={(1.5,0)}]
\node at (0,0) {$+~3~\times$};
\end{scope}

\begin{scope}
[shift={(1.25,-1)}]
\node at (0,0) {$+$};
\node at (0,-0.75) {$+$};
\node at (0,-1.5) {$+$};
\end{scope}

\begin{scope}
[shift={(2,0)}]
\draw (0,0.25)--(0.25,0);
\draw (0,-0.25)--(0.25,0);
\draw [fill=black](0.25,0) circle (0.05cm);
\draw (0.5,0) circle (0.25cm);
\draw [fill=black] (0.4,-0.15)--(0.6,-0.35);
\draw [fill=black] (0.4,-0.35)--(0.6,-0.15);
\draw [fill=black](0.75,0) circle (0.05cm);
\draw (0.75,0)--(1,0.25);
\draw [fill=black](0.5,0.25) circle (0.05cm);
\draw (0.5,0.25)--(0.75,0.5);
\draw (0.5,0.25)--(0.25,0.5);
\draw (0.75,0)--(1,-0.25);
\end{scope}

\begin{scope}
[shift={(2.4,-1)}]
\node at (0,0) {$ \frac{\lambda^3 \phi_0^6}{16}\tilde{\int} \frac{1}{\left(l^2+m^2\right)^4 } $}; 
\end{scope}

\begin{scope}
[shift={(2.25,-1.75)}]
\node at (0,0) {$ \frac{i\lambda^3 \phi_0^6}{3\cdot 2^9\pi^2 m^4}$}; 
\end{scope}

\begin{scope}
[shift={(2.25,-2.5)}]
\node at (0,0) {$ \frac{i\lambda^3 \phi_0^6}{3\cdot 2^9\pi^2 m^4}$}; 
\end{scope}

\begin{scope}
[shift={(4,0)}]
\node at (0,0) {$ +~~~~~\dots$};
\node at (0,-1) {$ +~~~~~\dots$};
\node at (0,-1.75) {$ +~~~~~\dots$};
\node at (0,-2.5) {$ +~~~~~\dots$};
\end{scope}

\begin{scope}
[shift={(1.75,-3.25)}]
\node at (-0.25,0.15) {$\underbrace{\qquad\qquad\qquad\qquad\qquad\qquad\qquad\qquad\qquad\qquad}$};
\node[font=\tiny] at (-0.25,-0.25) {$-\frac{im^2}{32\pi^2}\left[\left(
(1+\frac{\lambda \phi_0^2}{2m^2}\right)\ln{\left[1+\frac{\lambda \phi_0^2}{2m^2}\right]}-\frac{\lambda \phi_0^2}{2m^2}\right]$};
\end{scope}

\end{tikzpicture}
\end{center}

\caption{Double check: traditional ``divergent" computation of the tree level and one loop contributions to $\Gamma_\theta$. Classical fields $\phi_0$ on the external lines, quantum fields $h$ on the internal lines. $\theta$-vertex (with Feynman rule $-1$) denoted by little crosses. $\tilde{\int} \equiv \int\frac{d^4l}{(2\pi)^4}$. ``REG" denotes (dimensional) regularisation, ``REN" denotes renormalisation following a condition derived from eqs.~\ref{ren4v} $\frac{d^2 \Gamma_\theta}{d \phi_0^2} \Big|_{\phi_0=0}= -i$. Vacuum diagram omitted. }

\label{gtfig}

\end{figure}
%%%%%%%%%%%%%%%%%%%%%%%%%%%%%%%%%%%%%%%%%%

It is worth noting that compared to the original CS papers, and to our works \cite{Mooij:2021ojy, Mooij:2021lbc} as well, eq.~\ref{bl} is the first time that analyticity is used in the CS mechanism. Actually, that brings a conceptual advantage. In the computations of this paper (be it the conventional one, or the CS one), two renormalisation (boundary) conditions suffice to fix the two integration constants in eq.~\ref{finres}, or, equivalently, the two Lagrangian parameters $m$ and $\lambda$ that appear in the effective potential. Extending our discussion to include the effective action as well, a third condition sets the field parameter. Everything is in order. 
%However, in the CS approach used so far, one imposes one more condition. In the $n$-point computations that condition is put on $\Gamma^{(2)}_\theta$, here it could be set on $\Gamma_\theta$. The existence of such an extra condition poses a problem: at the end of the day, it cannot be that a new approach to renormalisation requires an extra condition. After all, the number of Lagrangian parameters is what it is. Although in both CS computations, such a additional condition appears quite natural\footnote{For correlation functions, one can use $\Gamma^{(2)}_\theta(k^2=0) = 1$. For the effective potential, the condition $d/ dm^2~d^2\Gamma_\theta / d\phi_0^2 =0|_{\phi_0=\phi_0^\star}$ does the job.\label{fn}}, we regard it as a conceptual advantage that in the computation presented above, analyticity by itself ensures that two renormalisation (boundary) conditions suffice to determine the effective potential.

Finally, it is straightforward to see how the finite CS method generalises to higher loop order. The function $\Gamma_{\theta\theta}$ is a convergent object and can therefore be computed up to all orders in $\hbar$ without ever running into divergences\footnote{Of course, one might run into subdivergences, but these can be safely dealt with through the standard skeleton expansion.}. For $\Gamma_{\theta}$ and $\Gamma$ itself the situation is precisely as for the $n$-point correlation functions themselves, treated already in \cite{Mooij:2021ojy, Mooij:2021lbc}. Once we have obtained their $\mathcal{O}\left(\hbar^n\right)$ contributions, these can be inserted at the rightmost term in the CS equations. Evaluating the appropriate number of derivatives on the resulting equation and using the boundary conditions at the left-hand side of the CS equations results in an equation for the $\mathcal{O}\left(\hbar^{n+1}\right)$ contributions. Meanwhile, imposing the boundary conditions yields the $\mathcal{O}\left(\hbar^{n+1}\right)$ contributions to the CS parameters $\{\beta,\gamma,\gamma_\theta\}$. This process can be repeated up to any desired order in $\hbar$.

\section{Two fields} 
\label{twof}
 {In this Section, we want to generalise the analysis of the previous Section to the case of two in\-ter\-acting scalar fields. The motivation for that is clear: it is the simplest context in which we can present the hierarchy arguments that this paper is meant to address. We spend some time on describing the vacuum configurations of the theory. After all, in the most elementary BSM scenarios we have our Higgs field sitting in its electroweak vacuum, while new heavy scalar fields are residing in their own vacuum.}

We study the theory described by the Lagrangian
\be
\L = -\frac{1}{2}\left(\d_\mu \phi\right)\left(\d^\mu \phi\right) -\frac{1}{2}\left(\d_\mu \Phi\right)\left(\d^\mu \Phi\right)-\frac{m^2}{2}\phi^2 - \frac{M^2}{2} \Phi^2 - \frac{\lambda_\phi}{4!}\phi^4 - \frac{\lambda_\Phi}{4!} \Phi^4 - \frac{\lambda_{\phi \Phi}}{4} \phi^2 \Phi^2\,. 
\label{twolag}
\ee
At the classical level, there are four different kinds of vacua now, as shown in Table~\ref{tst55}. Depending on the sign of their mass parameter, both fields can be either in the symmetric phase or in the broken phase. We will be mostly interested in the vacua of type D, where both fields have negative mass squared parameters and reside at nonzero field values.

\begin{table}[t!]\large
  \begin{center} 
    \begin{tabular}{l|l|l|l|l} 
   & $v^2$ & $V^2$ & $ \frac{\d^2 \Gamma_{\rm cl}}{\d\phi^2}\Big|_{\phi=v, \Phi=V}$ & $\frac{\d^2 \Gamma_{\rm cl}}{\d\Phi^2}\Big|_{\phi=v, \Phi=V}$\\
    \hline\\
   $A$ & $0$&$0$&$m^2$&$M^2$\\
   $B$ & $-\frac{6m^2}{\lambda_\phi}$ & $0$ & $-2m^2$ & $M^2 -\frac{3 m^2 \lambda_{\phi\Phi}}{\lambda_\phi}$\\
   $C$ & $0$ & $-\frac{6M^2}{\lambda_\Phi}$ & $m^2-\frac{3M^2\lambda_{\phi\Phi}}{\lambda_\Phi}$ & $-2M^2$\\
   $D$ & $\frac{6\left(m^2 \lambda_\Phi-3 M^2 \lambda_{\phi\Phi}\right)}{9\lambda_{\phi \Phi}^2-\lambda_\phi \lambda_\Phi}$ &$ \frac{6 \left(M^2\lambda_\phi-3 m^2\lambda_{\phi\Phi}\right)}{9\lambda_{\phi \Phi}^2-\lambda_\phi \lambda_\Phi}$& 
   $\frac{2\lambda_\phi\left(m^2 \lambda_\Phi-3M^2\lambda_{\phi \Phi}\right)}{9\lambda_{\phi \Phi}^2-\lambda_\phi \lambda_\Phi}$
   & $\frac{2\lambda_\Phi\left(M^2\lambda_\phi-3m^2\lambda_{\phi\Phi}\right)}{9\lambda_{\phi\Phi}^2-\lambda_\phi \lambda_\Phi}$
     \end{tabular}
      \end{center}
      \caption{Exploring the 9 extrema (1 described by situation $A$, 2 by both $B$ and $C$, and 4 by $D$) of the two-field classical (tree level) potential taken from eq.~\ref{twolag}. $v$ and $V$ denote the vevs of the fields $\phi$ and $\Phi$ respectively. $\Gamma_{\rm cl}$ is the classical potential defined by eq.~\ref{twolag}.} 
      \label{tst55}
\end{table}
      
At such type D vacua, the mass matrix can be easily depicted in terms of the vevs $v$ and $V$ of the fields $\phi$ and $\Phi$ respectively:
\be
\mathcal{M}^2 = \left(\begin{array}{ccc}
\lambda_\phi v^2/3 && \lambda_{\phi\Phi}V \cdot v\\
\lambda_{\phi\Phi} V \cdot v && \lambda_\Phi V^2/3
\end{array}\right)
\label{massmat}
\ee
with $v$ and $V$ the vevs appearing in the first columns in Table~\ref{tst55}.

It is now clear that one single tuning of the Lagrangian parameters
\be
m^2 \lambda_\Phi - 3M^2 \lambda_{\phi \Phi} \qquad \to \qquad 0 \label{tune}
\ee
can keep the vev $v$ of the field $\phi_0$ as small as we want.  {This condition implies that $m^2$ is large, $m^2\sim M^2$  (we take $\lambda_\phi \sim \lambda_{\phi\Phi} \sim \lambda_\Phi$ for power counting). Eq. \ref{tune} requires cancellations (fine-tunings) between different Lagrangian parameters $m^2$, $M^2$ and coupling constants. If the ``fundamental parameters'' of the theory are chosen to be the vevs $v$ and $V$, the hierarchy condition $v \ll V$ does not require any cancellations between large numbers. Meanwhile, the same tuning sends the smallest eigenvalue of the matrix (\ref{massmat}) - the physical mass $m_{\rm phys}$ of the lightest scalar to zero.  In the limit $v \ll V$ the $\phi$ and $\Phi$ are still the approximate true mass eigenstates of the theory.}  (Of course, tuning the quantity $M^2\lambda_\phi-3m^2\lambda_{\phi\Phi}$ leads to the same conclusions, but with the fields $\phi$ and $\Phi$ interchanged.)

Moving on to the quantum level, the definition of the two-field effective action is a straightforward generalisation of eq.~\ref{pos}:
\be
\Gamma_{\rm eff} = \sum_{n,N}  \frac{1}{(n+N)!}\int d^4x_1\dots d^4 x_{n+N}~\Gamma^{(n,N)}(x_1,\dots,x_{n+N})~\phi_0(x_1)\dots,\phi_0(x_{n})~ \Phi_0(x_{n+1})\dots,\Phi_0(x_{n+N}), \label{Geff2}
\ee
where $n$ denotes the number of $\phi$ fields and $N$ the number of $\Phi$ fields. In our case, our goal is again to compute the effective potential $\Gamma$ which (following the notation of eq.~\ref{derexp}) is related to the effective action $\Gamma_{\rm eff}$ through 
\be
\Gamma_{\rm eff} = -i\int d^4 x \left[ \Gamma(\phi_0,\Phi_0)+ \frac{1}{2}\left(\d_\mu \phi_0\right)^2 Z_\phi(\phi_0,\Phi_0) + \frac{1}{2}\left(\d_\mu \Phi_0\right)^2 Z_\Phi(\phi_0,\Phi_0)+\dots   \right].
\ee

\subsection{Standard computation}
\label{st1}
 {This Subsection contains a review of the standard approach to derive the two-field effective potential. It is much less explicit than its one-field counterpart, Subsection \ref{stapp}. We write down renormalisation conditions and give the final result in eq.~\ref{renres}.}
To not obscure our reasoning, we will avoid explicit (intermediate) expressions as much as we can. The effective action follows from summing all Feynman diagrams with external classical lines sticking out of purely quantum loops. Diagrams with $n$ light fields and $N$ heavy fields on the external lines give contributions of order $\phi_0^n\Phi_0^N$. Divergences are regulated, for example by dimensional regularisation. The instructions for their subtraction are encoded in renormalisation conditions. Omitting the constant term, we have 7 renormalisation conditions for fitting 7 Lagrangian parameters (two kinetic terms for the fields, two masses, and three four-point couplings). Restricting ourselves to the effective potential as we did before, the number of Lagrangian parameters to fit and the conditions to do so reduces to 5. In the spirit of eqs.~\ref{ren4v}, we choose
\ba
\frac{d^2 \Gamma} {d\phi_0^2}\Big|_{\phi_0=v,\Phi_0=V}= \frac{d^2 \Gamma_{\rm cl}}{d\phi_0^2}\Big|_{\phi_0=v,\Phi_0=V}\,,\qquad
\frac{d^2 \Gamma} {d\Phi_0^2}\Big|_{\phi_0=v,\Phi_0=V}= \frac{d^2 \Gamma_{\rm cl}}{d\Phi_0^2}\Big|_{\phi_0=v,\Phi_0=V}\,,\nn\\
i\cdot\frac{d^4 \Gamma} {d\phi_0^4}\Big|_{\phi_0=v,\Phi_0=V}= \lambda_\phi\,, \qquad
i\cdot\frac{d^4 \Gamma} {d\phi_0^2 d\Phi_0^2}\Big|_{\phi_0=v,\Phi_0=V}= \lambda_{\phi\Phi}\,,\qquad
i\cdot\frac{d^4\Gamma} {d\Phi_0^4}\Big|_{\phi_0=v,\Phi_0=V}= \lambda_\Phi\,. 
\label{s2a}
\ea
The right-hand sides of the first and second condition can be read off directly from Table \ref{tst55}. Again the philosophy is clear: we impose that at the vev, all quantum corrections to the second and fourth derivatives of the effective potential vanish.

Alternatively, one can choose to replace the conditions on the second derivatives with conditions on the first derivatives:
\ba
\frac{d \Gamma} {d\phi_0}\Big|_{\phi_0=v,\Phi_0=V}&=& 0,\nn\\
\frac{d \Gamma} {d\Phi_0}\Big|_{\phi_0=v,\Phi_0=V}&=& 0\,.
\label{s2b}
\ea
Note that this approach only works for vacua of type D where both fields are in the broken phase. As we have seen in the one field case, in the symmetric phase the first derivative of the potential vanishes anyway. Only in case D the third and fourth conditions are valid (``not-void") and all divergences can be removed.

Explicit expressions for the counter-terms required to meet these conditions are tedious and will be omitted. What matters for our argument is the observation that the counterterm for $\phi$'s mass (squared) parameter picks up contributions from diagrams with $\Phi$ running through the loops\footnote{Of course, there are also contributions $\delta M^2 \propto m^2$, but these are not relevant for the usual naturalness arguments.}.
\be
\delta m^2 \propto M^2.  \label{dmsq}
\ee
This observation is generic for all choices of renormalisation conditions and renormalisation points at which these conditions are to be evaluated. In the next section, where we will assume that $m \ll M$, it leads to the naturalness worries that this paper is meant to address. 

Going through all usual steps, one ends up with
\ba
\Gamma &=&  \frac{m^2}{2}\phi_0^2 + \frac{M^2}{2} \Phi_0^2 + \frac{\lambda_\phi}{4!}\phi_0^4 + \frac{\lambda_\Phi}{4!} \Phi_0^4 + \frac{\lambda_{\phi \Phi}}{4} \phi_0^2 \Phi_0^2\nn\\
&&   \qquad +\frac{\hbar}{64\pi^2}\Biggl(m_1^4(\phi_0,\Phi_0)
\ln{\left[\frac{m_1^2(\phi_0,\Phi_0)}{m^2} \right]}
+m_2^4(\phi_0,\Phi_0)\ln{\left[\frac{m_2^2(\phi_0,\Phi_0)}{M^2} \right]}\nn\\
%&& \qquad  \qquad +\frac{\hbar}{64\pi^2}\Biggl(\left(m^2 +\frac{\lambda_\phi}{2}\phi_0^2 + \frac{\lambda_{\phi\Phi}}{2}\Phi_0^2\right)^2\ln{\left[ 1+\frac{\lambda_\phi \phi_0^2 +\lambda_{\phi\Phi}\Phi_0^2}{2m^2} \right]}\nn\\
%&& \qquad \qquad \qquad \qquad+\left(M^2 +\frac{\lambda_\Phi}{2}\Phi_0^2 + \frac{\lambda_{\phi\Phi}}{2}\phi_0^2\right)^2\ln{\left[ 1+\frac{\lambda_\Phi \Phi_0^2 +\lambda_{\phi\Phi}\phi_0^2}{2M^2} \right]} \nn\\
&&  \qquad \qquad -\alpha_1\cdot m^2 \phi_0^2 - \alpha_2\cdot M^2 \Phi_0^2 - \alpha_3 \cdot \Phi_0^4 - \alpha_4 \cdot \phi_0^2\Phi_0^2- \alpha_5 \cdot \Phi_0^4
\Biggr).  \label{renres}
\ea
Here the coefficients $\{\alpha_1,\dots,\alpha_5\}$ are (long) functions of the Lagrangian parameters and $m_{1,2}^2(\phi_0,\Phi_0)$ are the eigenvalues of the mass matrix in the arbitrary background fields $\phi_0,\Phi_0$ that we will not bother to write out.  {We stress that these coefficients are all finite. They are the result of subtracting infinite and large counterterms from infinite and large counterterms. At the end of the day, however, the result in eq.~\ref{renres} gives only small corrections to the tree effective potential in the vicinity of the classical minimum ($\phi_0,\Phi_0$) thanks to the imposed boundary conditions (\ref{s2a}) or (\ref{s2b}). Corrections coming from the heavy sector are suppressed by powers of $v/V$. Our goal in the next subsection will be to arrive at the same result without running into infinite and large loop corrections that need to be eliminated by carefully picking infinite and large counterterms.}

\subsection{Convergent computation}
\label{conv1}
In our paper \cite{Mooij:2021lbc} we have given a detailed analysis of the two-field generalisation of the CS method for correlation functions. Here our task is to rewrite that to a two-field generalisation of the CS method for the effective action. We will go rather fast in this section since all details follow straightforwardly from the corresponding details described in \cite{Mooij:2021lbc}.

What is more, we stress that the goal of this section (and of the whole paper) is not to convince the reader that from now on actual computations should be done by the manifestly convergent CS approach. The expressions appearing in this Subsection make clear that for actual computations, the standard textbook ``divergent" framework is much more user-friendly. Our goal is only to show the {\it existence} of a valid framework for an alternative computation of the (two-field) effective potential.  {The central equations in this Subsection are \ref{clight}, \ref{cheavy} and \ref{matr2}. Their relevance lies in their mere existence, not in their actual shape and structure.}

With two fields and two masses on the scene, we have two kinds of $\theta$-operations. The definitions in eq.~\ref{thetdef2} are generalised to:
\ba
\tilde{\Gamma}_{\theta,m} \equiv -i \times \frac{d}{d m_0^2} \tilde{\Gamma}\,, \qquad
\tilde{\Gamma}_{\theta,M} \equiv -i \times \frac{d}{d M_0^2} \tilde{\Gamma}
\label{gt2}
\ea
and then
\ba
\tilde{\Gamma}_{\theta \theta,mm} \equiv -i \times \frac{d}{d m_0^2} \tilde{\Gamma}_{\theta,m}\,, \qquad
\tilde{\Gamma}_{\theta\theta,Mm} \equiv -i \times \frac{d}{d M_0^2} \tilde{\Gamma}_{\theta,m}\,, \nn\\
\tilde{\Gamma}_{\theta \theta,mM} \equiv -i \times \frac{d}{d m_0^2} \tilde{\Gamma}_{\theta,M}\,, \qquad
\tilde{\Gamma}_{\theta\theta,MM} \equiv -i \times \frac{d}{d M_0^2} \tilde{\Gamma}_{\theta,M}\,,
\label{gtt2}
\ea
where of course $\tilde{\Gamma}_{\theta,Mm} = \tilde{\Gamma}_{\theta,mM}$. Again the tildes indicate bare quantities. As in the one field case the equations derived from these definitions (to be presented below) do not contain bare objects. 

The corresponding CS equations (two-field generalisations of eqs.~\ref{effcs}) show how the light and heavy sectors are intertwined. The two $\Gamma_\theta$s defined in eqs.~\ref{gt2} can be united in a two-dimensional vector, while the four $\Gamma_{\theta\theta}$s form a two-by-two matrix. Written out in components, the first CS equation generalises to
\ba
2i\left[m^2\left(1+\gamma_{\phi,m}\right)\Gamma_{\theta,m} + M^2 \gamma_{\Phi,m}\Gamma_{\theta,M}\right] &=& \Bigg[2m^2\frac{\d}{\d m^2}+\sum_i\beta_{\lambda_i,m}\frac{\d}{\d \lambda_i}\nn\\
&&   \qquad\qquad+ \gamma_{\phi,m}\cdot\phi\frac{\d}{\d \phi}+\gamma_{\Phi,m}\cdot \Phi\frac{\d}{\d \Phi}\Bigg]\Gamma \label{clight}
\ea
and
\ba
2i\left[M^2\left(1+\gamma_{\Phi,M}\right)\Gamma_{\theta,M} + m^2 \gamma_{\phi,M}\Gamma_{\theta,m}\right] &=& \Biggl[2M^2\frac{\d}{\d M^2}+\sum_i\beta_{\lambda_i,M}\frac{\d}{\d \lambda_i}\nn\\
&&   \qquad\qquad+ \gamma_{\phi,M}\cdot\phi\frac{\d}{\d \phi}+\gamma_{\Phi,M}\cdot \Phi\frac{\d}{\d \Phi}\Bigg]\Gamma  ~.\label{cheavy}
\ea
Here the sum over $\lambda_i$ runs over the three four-point couplings $\{\lambda_\phi,\lambda_\Phi, \lambda_{\phi\Phi}\}$. 

The second CS equation reads in full two-by-two glory
\ba
 i\cdot &&\left(
\begin{array}{ccc}
1+\gamma_{\phi,m} &  \frac{M^2}{m^2} \gamma_{\Phi,m} \\
 \frac{m^2}{M^2} \gamma_{\phi,M} & 1+\gamma_{\Phi,M}
\end{array}
\right) \times  \left(
\begin{array}{ccc}
\Gamma_{\theta\theta,mm} & \Gamma_{\theta\theta,mM}\\
\Gamma_{\theta \theta, Mm}&\Gamma_{\theta \theta,MM}
\end{array}
\right)\nn\\
&& \qquad \qquad=
\Biggl[
\left(\begin{array}{ccc}
\d / \d m^2 \\ \d/\d M^2\end{array}\right)
+\sum_i \left(\begin{array}{ccc}
\beta_{\lambda_i,m}/2m^2 \\ \beta_{\lambda_i,M}/2M^2
\end{array}\right)
\frac{\d}{\d \lambda_i}\nn\\
&& \qquad \qquad \qquad\qquad
+\frac{1}{2}\left(\begin{array}{ccc}
\gamma_{\phi,m}/m^2 \\ \gamma_{\phi,M}/M^2
\end{array}\right)\phi\frac{\d}{\d \phi}
+\frac{1}{2}\left(\begin{array}{ccc}
\gamma_{\Phi,m}/m^2 \\ \gamma_{\Phi,M}/M^2
\end{array}\right) \Phi\frac{\d}{\d \Phi}
\Biggr]
\left(\begin{array}{ccc}
\Gamma_{\theta,m} & \Gamma_{\theta,M}\end{array}\right)\nn\\
&& \qquad \qquad\qquad+
\frac{1}{2m^2} 
\times
 \left(\begin{array}{ccc}
\gamma_{\theta,mmm} & \gamma_{\theta,Mmm} \\
\gamma_{\theta,mmM} & \gamma_{\theta,MmM}
\end{array}\right) 
 ~\Gamma_{\theta,m}\nn\\
 &&\qquad\qquad \qquad
+
\frac{1}{2M^2}
\times
\left(\begin{array}{ccc}
\gamma_{\theta,mMm} & \gamma_{\theta,MMm} \\
\gamma_{\theta,mMM} & \gamma_{\theta,MMM}
\end{array}\right)
 ~\Gamma_{\theta,M}~. \label{matr2}
\ea
Note that the auxiliary CS parameters $\{\beta,\gamma,\gamma_\theta\}$ in the one field case have now generalised to 6 $\beta$s (related to two different derivatives on three couplings), 4 $\gamma$s (related to two different derivatives on two fields) and 8 $\gamma_{\theta}$s (indirectly related to two different derivatives on two mass parameters related to two fields).

\begin{figure}[!b]
 \centering
    \includegraphics[width=0.6\textwidth]{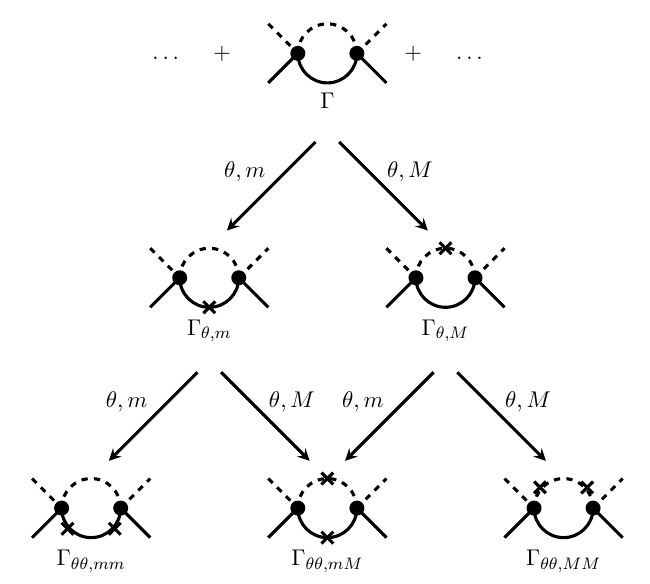}
\caption{Two-field generalisation of the left-hand side of Fig. \ref{finfig}. Straight lines are for $\phi$, dashed lines are for $\Phi$. At the top, we have an (infinite) sum of diagrams that together constitute the full effective action. For the sake of readability, we only depict the effect of $\theta$-operations on one of these diagrams. As always, external lines are classical fields, internal lines are quantum fields.
}
\label{v2esimp}
\end{figure}

In Fig.~\ref{v2esimp} the structure of the two-field CS approach is outlined. For notational simplicity, we have only graphically depicted the $\theta$-operations on one of the contributing diagrams, but the discussion is completely generic. (And again, given the $\Gamma_{\theta\theta}$s and the boundary conditions there is strictly speaking no need to think about Feynman diagrams whatsoever.)

At the bottom line, we have three kinds of $\Gamma_{\theta\theta}$s. These can all be obtained by a divergence-free computation. (That is, if we decide again to leave out the constant contribution to the effective potential.) 

Inserting these $\Gamma_{\theta\theta}$s in the second CS equation we obtain the two $\Gamma_\theta$s. Next, the first CS equation brings us to $\Gamma$.  This process yields both mass derivatives (with respect to $m^2$ and $M^2$) of the five objects in eq.~\ref{s2a}, plus the two mass derivatives of the two $\Gamma_\theta$s. In the process of deriving them, one uses the CS parameters. Six of them (the $\beta$s) are involved in the computation of the two mass derivatives of the three quartic terms in $\Gamma$. Out of the other eight ones (the $\gamma_\theta$s), four of them are involved in the computation of the mass derivatives of the two mass terms in $\Gamma$. The last four are involved in the computation of the two mass derivatives of the quadratic terms in $\Gamma_\theta$ (using the second CS equation). All in all, to derive the full effective action we have 14 objects to fix, for which we use 14 renormalisation conditions and which yield the 14 CS parameters\footnote{Generalising the discussion to the effective action, we have four more objects (both mass derivatives on both kinetic terms) that can be fixed by the four $\gamma$ parameters $\{\gamma_{\phi,m},\gamma_{\Phi,m}, \gamma_{\phi,M}, \gamma_{\Phi,M}\}$.} We have all the ingredients to arrive at the result in eq.~\ref{renres} without meeting any intermediate divergence.
%%%%%%%%%%%%%%%%%%%%%%
%%%%%%%%%%%%%%%%%%%%%%%%%%
\section{Quantum stability of the two-field effective potential} 
\label{stable}
The theory of eq.~\ref{twolag} treats $\phi$ and $\Phi$ on equal footing. However, for our purposes, we will now assume that  {$v\ll V$}. We also assume that the three four-point couplings $\{\lambda_\phi, \lambda_{\phi\Phi}, \lambda_\Phi\}$ are of the same order (or that at least the hierarchy between them is much smaller than the hierarchy between the two mass parameters).

First of all, the interaction between $\phi$ and $\Phi$ gives off-diagonal elements to their mass matrix. Therefore, even before taking quantum corrections into account, the fields $\phi$ and $\Phi$ are not canonically normalised. However, in the situation  {$v \ll V$} there will be a ``light" canonical field $\phi_{\rm light}$ which almost coincides with $\phi$, and a ``heavy" canonical field $\phi_{\rm heavy}$ which almost coincides with $\Phi$. (Here ``almost" means: up to corrections proportional to  {$v/V$}.) To not lose ourselves in non-essential details, we will omit these redefinitions. Sloppily but safely we will refer to $\phi$ as the light field and to $\Phi$ as the heavy field.

To make our point, we consider two-field vacua of kind D. We have seen that at tree level, one single fine-tuning of the factor $\left(m^2 \lambda_\Phi-3M^2 \lambda_{\phi \Phi}\right)$ keeps both the vev of the light field and its mass small (i.e., free of contributions of order  {$V^2$}).

Next, we turn to quantum corrections. In the standard approach, these follow from an expansion in Feynman diagrams that contains both infinite and ``large" (order $M^2$) terms\footnote{Most notably, the bare effective potential contains terms of type $M^4 \ln{\left[1+\frac{\phi_0^2}{M^2}\right]}$ which around $\phi$'s vev produce contributions proportional to $M^2\phi_0^2$.}. In particular, we have diagrams of mass dimension two with two external classical lines which correct the second derivatives (the mass terms) of the effective action. To eliminate these divergences, we choose to work with the renormalisation scheme defined in eq.~\ref{s2a}. By construction, at the vevs the masses of $\phi$ and $\Phi$ are equal to the tree-level masses, just because we impose that at the vev the quantum corrections to the second derivatives of the effective action disappear. However, it is clear that to achieve this, the standard approach requires a delicate cancellation between both infinite and ``large" (order  {$V^2$}) loop corrections and counter-terms. 

For a typical GUT theory with  {$v\simeq 10^2~ {\rm GeV}$} and  {$V\simeq 10^{14}~ {\rm GeV}$}, the necessary tuning between the mass (squared) parameters $m^2$ and $M^2$ would be of order $10^{24}$, at least. It seems that to fulfil the renormalisation conditions, a new tuning is needed to restore the small (order  {$v$}) values for the vev of the field $\phi$. Even worse, at every new order in perturbation theory, a new tuning between the Lagrangian parameters is required. This is how the technical hierarchy problem appears in the computation of the effective action. The physics of the light field seems to be unstable against quantum corrections. Without careful fine-tuning that needs to be ``refreshed" at every next loop order, the coupling between $\phi$ and $\Phi$ shifts the mass of $\phi$ to values of order  {$V$}. 

On the other hand, applying the finite CS formalism results in the same effective potential given in eq.~\ref{renres} that respects the boundary conditions of eqs.~\ref{s2a}. Now quantum corrections follow from a diagrammatic expansion of the manifestly convergent object $\Gamma_{\theta\theta}$. Since $\Gamma_{\theta\theta}$ has mass dimension minus two, by construction the finite computation cannot contain any contributions and cancellations between divergent or large (order  {$V^2$}) quantities.

A last point that one might worry about is what quantum corrections do to the vev of the light field. After fixing $\Gamma$'s second and fourth derivatives we have no more handles left to tune, we can only compute the resulting vevs and hope that $v$ has not shifted to large values of order $M$. In other words, even if heavy exotic physics does not shift the mass of the light field, it might have shifted its vev, and the technical hierarchy problem reappears. This problem applies both to the standard method and to the CS method: at the end of the day, both have produced the same result. 

In the particular case of the theory described by eq.~\ref{twolag}, it is straightforward to see that the classical vev does not suffer enormous shifts when quantum corrections are taken into account. Since the resulting effective action in eq.~\ref{renres} is quadratic in $\phi_0$ and $\Phi_0$, its first derivative with respect to $\phi_0$ must be proportional to $\phi_0$. At the classical vev, it then becomes proportional to that same classical vev. With that, we understand that the same tree level tuning in eq.~\ref{tune} keeps the ``quantum corrected" vev at the classical vev. Moreover, three of the four elements in the mass matrix of eq.~\ref{massmat} are still proportional to that same factor. That ensures that the renormalisation scheme given in eqs.~\ref{s2a} produces an effective action in which one single tuning shields the physics of the light field $\phi$ from the new physics of the heavy field $\Phi$. Around the classical vevs, the renormalised theory does not show any imprint of the possible existence of new exotic physics living at a hypothetical scale  {$V \gg v$}.

However, what if we wish to go beyond the toy model of eq.~\ref{twolag} and consider more realistic (GUT) settings? The reason that the small vev does not shift in our toy model was a discrete symmetry $\phi \leftrightarrow -\phi$, forbidding the terms proportional to $\phi\Phi$ and $\phi\Phi^3$ in the effective potential\footnote{ {If the $\phi\Phi$, $\phi\Phi^3$, $\phi^3\Phi$ terms are introduced, the arguments about the hierarchy problem remain in force: in the CS approach with appropriate boundary conditions no tuning is necessary to keep the light scalar mass small in comparison with the heavy mass. However, the relationship between the vevs of the fields and the physical masses of the particles is lost - small vev of the light field does not automatically imply its small mass and vice-versa.}}. In GUTs like SU(5), similar types of terms are forbidden by the fact that the two breaking stages of the unifying group are carried out by different multiplets. When we choose the boundary (renormalisation) conditions that forbid corrections to the vev (to the mass), we get for free that the mass (the vev) does not move to much larger values. 

In summary, we have shown how all possible fine-tunings in the computation of the effective potential can be circumvented by applying the finite CS method. Also, we have made clear that in general the masses and vevs of the light field can be protected. Therefore, we can only conclude that the ``hierarchy worries" suggested by the traditional approach are nothing but an unphysical artefact of that approach. In other words: there is no reason to worry about alleged cancellations between unphysical Lagrangian parameters.

\section{Conclusions} 
\label{conclu}
In this paper, we have presented a manifestly finite approach to the computation of the effective action. All intermediate divergences can be circumvented. We have explicitly shown how the method works for the simple case of $\lambda \phi^4$ theory. We have then shown, by counting the number of parameters, that there is no fundamental obstruction in the two-field (or, $n$-field) generalisation of this computation. By no means do we suggest that one should actually {\it use} this system for explicit computations. In practice, the standard ``divergent" computation of the effective action (and potential) is much more direct and user-friendly. We just want to point out that to compute the effective action it is not necessary to resort to large (formally infinite) cancellations between bare loop corrections and ad hoc introduced artificial counter-terms.

Our motivation for this exercise lies in the so-called ``technical hierarchy problem." It is often assumed (see for example \cite{Gildener:1976ai}) that in theories with a large hierarchy between (mass) scales careful tunings are necessary to shield low-energy physics from the influence of new exotic physics going on at much higher scales. In the particular case of the effective action/potential, it seems that at every order in perturbation theory, one needs a new tuning to prevent the vevs and masses of the light theory from running away to (much higher) energy scales associated with new physics. We have shown that the CS approach that we present in this paper does not need delicate ``re-fine-tunings" to ``keep light physics light". From there, we argue that the ``hierarchy worries" about the alleged quantum-mechanical instability of the effective action/potential are method-dependent and therefore unphysical. In perfect agreement with the decoupling theorem, ``accessible physics" at the electroweak scale never feels possible new physics at GUT scales or beyond.

\section*{Acknowledgements}
This work was supported by the Generalitat Valenciana grant PROMETEO/2021/083. SM wishes to thank the Nikhef Theory Group for their everlasting hospitality; part of this work was done during such a stay in Amsterdam.

\begin{appendix}
\section{Inclusion of cosmological constant} \label{cc}

In QFT without gravity, there is no physical meaning in the absolute value (that is, the constant, $\phi_0$-independent part) of the effective potential. However, a priori there seems to be no obstruction in applying the CS program to the computation of that first, constant term that shows up in the effective potential. Our paper \cite{Mooij:2021ojy} contains a short section on this constant term. In this Appendix, we extend and clarify the discussion there. For simplicity, we restrict ourselves to the case of a single scalar field. Also, we will mostly work ``top-down": we merely want to show that the CS approach to computing the effective potential including its constant term is completely well-defined and consistent. With that, it is straightforward to do actual computations as we did in Section \ref{fin}.

\begin{figure}[!b]
 \centering
    \includegraphics[width=0.7\textwidth]{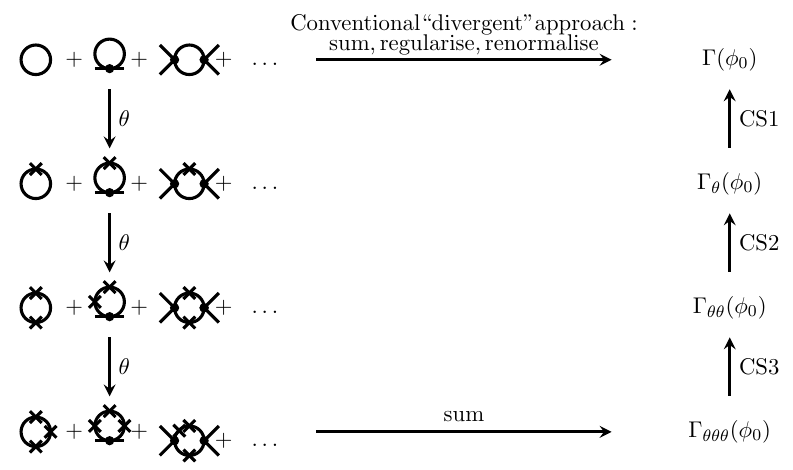}
\caption{Three-step generalisation of Fig. \ref{finfig}}.
\label{finfig2}
\end{figure}

The constant part of the effective action follows from the first diagram appearing in the diagrammatic expansion of the effective action: the vacuum bubble diagram. As it has mass dimension four, it is clear that we will need three $\theta$-operations to compute the full effective action. Consequently, we need one more CS equation. That equation should instruct us how to get $\Gamma_{\theta\theta}$ out of the manifestly convergent object $\Gamma_{\theta\theta\theta}$. See Figure \ref{finfig2} for a sketch of this extended CS program.

Going through the derivation of the first two CS equations in Callan's original notes \cite{Callan:1975vs}, it seems that the third equation should be \cite{Mooij:2021ojy}
\be
2im^2~(1+\gamma) \cdot \Gamma_{\theta\theta\theta} =\left(2m^2~\frac{\d}{\d m^2} +\beta~\frac{\d}{\d \lambda} +\gamma \cdot \phi_0\frac{\d}{\d \phi_0}+2\gamma_\theta\right)\Gamma_{\theta\theta}. \label{CS3}
\ee
On the left-hand side, we find the object $\Gamma_{\theta\theta\theta}$, which has mass dimension minus two. As such, it can be computed from a series of convergent diagrams, just as we want. See again Figure \ref{finfig2}. One readily finds
\be
\Gamma_{\theta\theta\theta}=\frac{i\hbar}{32\pi^2 m^2}\left[1-\frac{\lambda\phi_0^2}{2m^2}+\frac{\lambda^2 \phi_0^4}{4m^4}+\dots\right] = \frac{i\hbar}{32\pi^2}\cdot\frac{1}{m^2+\lambda\phi_0^2/2}~.\label{vttt2}
\ee
To show the consistency of this extended CS mechanism, we can take the standard result (computed by the traditional approach), throw it through the three CS equations beginning from the right top of Fig.~\ref{finfig2}, and see whether we end up with the above result for $\Gamma_{\theta\theta\theta}$. 

However, in doing so we stumble on a problem. We begin with the symmetric case. Throwing the result in eqs.~\ref{ve} and \ref{vecoeff} through the three CS equations given in eqs.~\ref{effcs} and \ref{CS3}, we find
\be
\Gamma_{\theta\theta\theta}=-\frac{i\hbar}{32\pi^2 m^2}\cdot \frac{\lambda\phi_0^2}{2m^2} \left[1-\frac{\lambda\phi_0^2}{2m^2}+\frac{\lambda^2 \phi_0^4}{4m^4}+\dots\right].\label{trubbels}
\ee
Clearly, we have lost the first term in the expansion in eq.~\ref{vttt2}. 

The solution to this paradox is to introduce a constant term (say, $\Lambda$) to our Lagrangian, and treat that as a proper parameter in the CS formalism. That is, we begin from the classical action
\be
\Gamma_{\rm cl}= -i\cdot \int d^4x \left[\Lambda +\frac{1}{2}\left(\d_\mu \phi_0\right)^2+\frac{m^2}{2}\phi_0^2 +\frac{\lambda}{4!}\phi_0^4 \right].
\ee
We renormalise the effective potential by imposing
\ba
\Gamma \Big|_{\phi_0=0} &=& \Lambda\nn\\
\frac{d^2 \Gamma}{d \phi_0^2} \Big|_{\phi_0=0} &=& m^2\nn\\
\frac{d^4 \Gamma}{d \phi_0^4} \Big|_{\phi_0=0} &=& \lambda.
\ea
That is going to give the result given in eqs.~\ref{ve} and \ref{vecoeff}, augmented by a cosmological constant $\Lambda$:
\ba
i\cdot \Gamma_{\rm eff} &=&   \int d^4 x \Biggl[\Lambda+\frac{1}{2}\left(\d_\mu \phi_0\right)^2 +\frac{m^2}{2}\phi_0^2 +\frac{\lambda}{4!}\phi_0^4\nn\\
&& \qquad \qquad +\frac{\hbar}{64\pi^2}\left(\left(m^2+\lambda\phi_0^2/2\right)^2\ln{\left[ 1+\frac{\lambda\phi_0^2/2}{m^2} \right]}-\alpha_0\cdot m^4 - \alpha_1 \cdot m^2 \phi_0^2 - \alpha_2 \cdot \phi_0^4
\right)\Biggr],\nn\\\label{casicasi}
\ea
with
\be
\alpha_0 = 0\,,\qquad
\alpha_1= \frac{\lambda}{2}\,, \qquad \alpha_2 = \frac{3\lambda^2}{8}.
\ee
Meanwhile, the CS equations are generalised to
\ba
2 im^2 ~(1+\gamma) ~\Gamma_\theta&=&  \left(2m^2 ~\frac{\d}{\d m^2} +\beta~\frac{\d}{\d \lambda}    + \gamma \cdot \phi_0\frac{\d}{\d \phi_0}+\gamma_\Lambda ~\frac{\d}{\d \Lambda}\right) \Gamma,\nn\\
2im^2~(1+\gamma) ~   \Gamma_{\theta\theta} &=& \left(2m^2~\frac{\d}{\d m^2} +\beta~\frac{\d}{\d \lambda} +\gamma \cdot \phi_0\frac{\d}{\d \phi_0}+\gamma_\theta+\gamma_\Lambda~ \frac{\d}{\d \Lambda}\right)\Gamma_\theta,\nn\\
2im^2~(1+\gamma) \cdot \Gamma_{\theta\theta\theta} &=&\left(2m^2~\frac{\d}{\d m^2} +\beta~\frac{\d}{\d \lambda} +\gamma \cdot \phi_0\frac{\d}{\d \phi_0}+2\gamma_\theta+\gamma_\Lambda~ \frac{\d}{\d \Lambda}\right)\Gamma_{\theta\theta}. \label{3css}
\ea
In line with the CS logic so far, the appearance of the new parameter $\Lambda$ in the theory results in a new CS parameter $\gamma_\Lambda$. As opposed to the other CS parameters, it is dimensionful. As a general Ansatz, we take for its one-loop description
\be
\gamma_\Lambda = \Lambda\cdot\left(c_1 + \frac{\hbar}{32\pi^2}\cdot \alpha_1 + \O\left(\hbar^2\right)\right) + m^4 \cdot\left(c_2 + \frac{\hbar}{32\pi^2}\cdot \alpha_2 + \O\left(\hbar^2\right)\right).
\ee
By a bottom-up computation, we find that taking
\ba
c_1 &=& 4+\O\left(\lambda\right), \qquad \alpha_1=-4\lambda + \O\left(\lambda^2\right)\nn\\
c_2 &=& \O\left(\lambda^2\right), \qquad \alpha_2 = 1+\O\left(\lambda^2\right)
\ea
does the job. We conclude that with the inclusion of the constant term $\Lambda$ and its associated CS parameter $\gamma_\Lambda$, the constant contribution that was missing in eq.~\ref{trubbels} reappears. With that, we have shown that indeed the constant contribution to the effective action can be accounted for by a slight generalisation of the Callan-Symanzik method.

Finally, we mention that in the SSB case treated in the body of the paper, the story is precisely the same. In the traditional way one finds that the renormalisation conditions imposed at the theory's vev
\ba
\Gamma \Big|_{\phi_0=v} &=& \Lambda-\frac{3m^4}{2\lambda}\,,\nn\\
\frac{d^2 \Gamma}{d \phi_0^2} \Big|_{\phi_0=v} &=& -2m^2\,,\nn\\
\frac{d^4 \Gamma}{d \phi_0^4} \Big|_{\phi_0=v} &=& \lambda,
\ea
again lead to the result of eq.~\ref{casicasi}, but now with (a direct generalisation of the result in eq.~\ref{brocof})
\be
\alpha_0=\frac{285}{2}+i\pi+\ln{2} \,, \qquad \qquad
\alpha_1 = \lambda\left(\frac{61}{2}+i\pi+\ln{2}\right)\,,  \qquad \alpha_2 = \frac{\lambda^2}{4}\left(\frac{9}{2}+i\pi+\ln{2}\right)\,. 
\ee
Similarly, imposing
\ba
\Gamma \Big|_{\phi_0=v} &=& \Lambda-\frac{3m^4}{2\lambda}\,,\nn\\
\frac{d \Gamma}{d \phi_0} \Big|_{\phi_0=v} &=& 0\,,\nn\\
\frac{d^4 \Gamma}{d \phi_0^4} \Big|_{\phi_0=v} &=& \lambda,
\ea
leads to eq.~\ref{casicasi} with
\be
\alpha_0=\frac{69}{2}+i\pi+\ln{2} \,, \qquad \qquad
\alpha_1 = \lambda\left(\frac{25}{2}+i\pi+\ln{2}\right)\,,  \qquad \alpha_2 = \frac{\lambda^2}{4}\left(\frac{9}{2}+i\pi+\ln{2}\right)\,. 
\ee
In both cases we find that same three CS equations in eqs.~\ref{3css} consistently connect these two results to the result for $\Gamma_{\theta\theta\theta}$ given in eq.~\ref{vttt2}.

%%%%%%%%%%%%%%%%%%%%%%
%%%%%%%%%%%%%%%%%%%%%%%%%%

\end{appendix}

\newpage
%\bibliographystyle{utphys}
%\bibliographystyle{JHEP}
%\bibliography{Potential}

\providecommand{\href}[2]{#2}\begingroup\raggedright\endgroup

\end{document}